\documentclass[a4paper]{article}

\usepackage{amsmath,amssymb,graphicx}
\usepackage{color}
\usepackage{siunitx}
\begin{document}

\title{Gap solitons and symmetry breaking in parity-time symmetric microring CROWs}
\author{Pedro Chamorro-Posada\\
Dpto. de Teor\'{\i}a de la Se\~nal y Comunicaciones \\
e Ingenier\'{\i}a Telem\'atica,\\ 
Universidad de Valladolid, ETSI Telecomunicaci\'on,\\
 Paseo Bel\'en 15, 47011 Valladolid, Spain}

\date{\today}
\maketitle

\begin{abstract}
The propagation properties of optical fields in linear and nonlinear parity-time symmetric microring coupled resonator optical waveguides are studied.  The effects described include the existence of symmetry breaking thresholds, the propagation of gap solitons in nonlinear transmission lines and the existence of quasi stable propagation regimes outside the broken symmetry regions.
\end{abstract}

\section{Introduction}

A non-Hermitian Hamiltonian can have an entirely real spectrum provided that parity-time ($\cal{PT}$) symmetry is preserved \cite{bender}.  A necessary condition for this \cite{bender2007} is that the complex potential satisfies $V(x)=V^*(-x)$.  These systems are typically characterized by a non-Hermiticity parameter that determines the existence of a symmetry breaking transition from real to complex eigenvalues  at a certain threshold. 

$\cal{PT}$ symmetry in the context of optical systems \cite{ganainy,makris,guo,longhi2010,peng,regensburger,ruter} has attracted considerable attention.  In optics,  parity-time symmetry is associated to the fact that the complex refractive index $n(\mathbf{r})$ satisfies the condition $ n^*(\mathbf{r})= n(-\mathbf{r})$.  This has been experimentally tested in various works \cite{guo,ruter,peng,regensburger,feng}. $\cal{PT}$ symmetric optical systems have opened up a series of new phenomena with potential applications, such as nonreciprocal optical transmission \cite{feng,feng2013,sukho,rame,peng,bender2013}, unidirectional invisibility \cite{lin,longhi}, loss-induced transparency \cite{guo} or coherent perfect absorption (CPA)\cite{chong,zhang}.

Soliton propagation in $\cal{PT}$ symmetric potentials periodic along the propagation coordinate were considered in \cite{miri} and \cite{liu} for a Kerr and a competing cubic-quintic nonlinearity, respectively.  Parity-time symmetric coupled microresonator have been recently addressed in \cite{peng, bender2013}.  Quite remarkably, the use of resonant structures has permitted to achieve a $\SI{1}{\micro\watt}$ threshold for the observation of symmetry-breaking non-reciprocal light transmission.  
Discrete solitons in $\cal{PT}$-symmetric nonlinear magnetic metamaterial consisting of split-ring dimers with gain and loss have also been recently considered in \cite{lazarides,wang}.

In this work, the nonlinear propagation in parity-time symmetric microring coupled resonator optical waveguides (CROWs)  is addressed.  This extends, by including the effect of periodic gain and loss, previous work \cite{jopt} on passive nonlinear microring CROW transmission systems.  In the tightly coupled regime, the system is closely related to a nonlinear grating \cite{jopt}.  Therefore, it is expected that these structures will find applications similar to those of $\cal{PT}$ symmetric Bragg gratings, such as unidirectional invisibility or CPA.  The work presented here is related with Refs. \cite{miri,peng,bender2013}, but also with the parity-time synthetic photonic lattices of \cite{regensburger} implemented using  coupled fiber loops.  There, the physical layout corresponds to a doubly discrete time-time dynamics, whereas their extended fiber network is a space-space discrete dynamical system analogue.  The microring CROW, on the other hand, has associated a space-time discrete evolution.

CROW transmission media \cite{Yariv99} have been proposed for the implementation of optical filters \cite{chamorro11} or the realization of fast and slow wave structures \cite{boyd,fraile,ol2009}. One particular benefit of slow-wave optical systems is the enhancement of the nonlinear optical response \cite{chen,melloni}.  Similarly to the case of nonlinear Bragg gratings \cite{parini07}, there are several propagation phenomena associated to nonlinear CROWs including multistability \cite{dumeige}, self-pulsation \cite{grigoriev2011,maes2009}, chaos \cite{grigoriev2011}, modulational instability \cite{huang2009} and gap solitons \cite{christo}.  The dynamics of nonlinear CROWs had been previously addressed using coupled mode theory for short chains \cite{maes2009,grigoriev2011} and discrete nonlinear Schr\"odinger models for long structures \cite{huang2009,christo}.  In \cite{jopt}, a fully discrete model was introduced and it was used in the analysis of the behavior of optical signals in microring CROWs.  This model is a generalization of the Ikeda equations for the single nonlinear ring resonator \cite{ikeda}.    

The propagation properties in nonlinear lossless structures are first reviewed.  Then, nonlinear microring CROW transmission lines with periodic gain and loss are considered.  When the gain-loss period coincides with that of the microring structure, no symmetry breaking transition is observed.  In this case, the main effect of the periodic gain-loss variation is to modify the effective nonlinearity for a given input power.  When the gain-loss period doubles the microring structure pitch, there exist two well-defined parameter regions bounded by two symmetry breaking transitions.  Outside the broken symmetry regions, the solutions are shown to be also unstable almost everywhere, but numerical simulations demonstrate the quasi stable propagation of optical signals in certain parameter regimes.  

\section{Lossless structures}

\begin{figure}
\centerline{\includegraphics[width=.9\columnwidth]{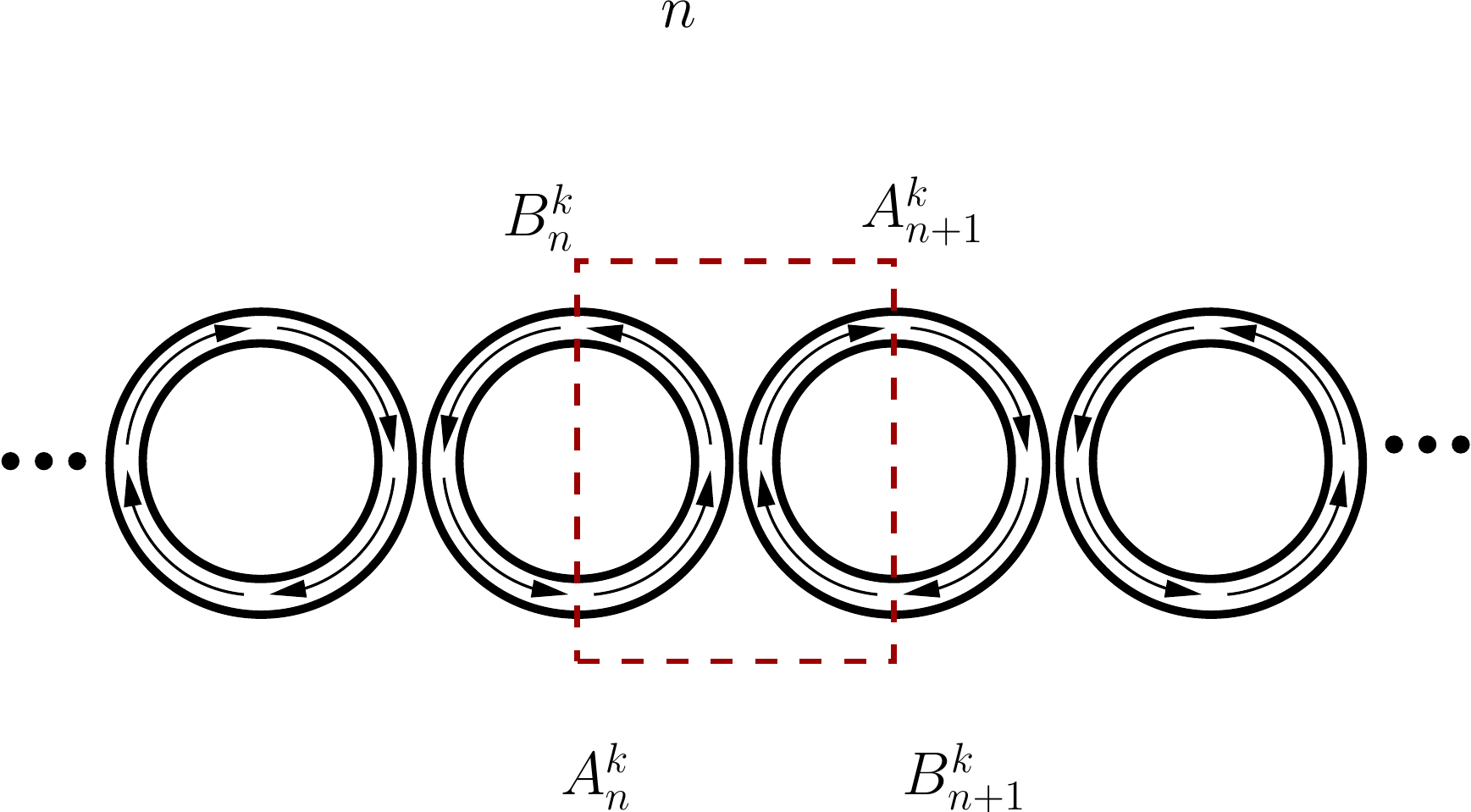}}
\caption{microring CROW structure.} \label{fig::crow0}
\end{figure}

The properties of passive lossless nonlinear microring CROW structures were studied in \cite{jopt}.  The system geometry is shown in Figure \ref{fig::crow0}. Evanescent coupling between sections is assumed to be precisely localized at a point and described by the real parameters $\theta$ and $\rho$, $\theta^2+\rho^2=1$ \cite{jopt}.  


  The propagation length across one section is $L=\pi R$, where $R$ is the resonator radius.  A convenient space-time discretization is obtained by setting the sampling period as the one-section propagation delay $\tau=L/v_g$, where $v_g$ is the group velocity.  Discrete time $k$ will correspond to $t_k=k\tau$ and discrete space $n$ will correspond to the $n$th CROW section, as shown in Figure \ref{fig::crow0}.  

A band-pass propagating optical signal with carrier frequency $\omega$ is assumed.   The signal transmission across the cell length $L$ has then an associated linear phase shift of $\exp(-j\Omega)$, where $\Omega=\omega \tau$.  The complex field envelopes of the forward and backward waves in the structure at discrete position $n$ and time $k$ are $A_n^k=A_n(t_k)$ and $B_n^k=B_n(t_k)$, respectively.  The effect of the Kerr nonlinearity is computed along sections of length $d=L/2$. The resulting nonlinear phase shift is $\exp\left(-j\Gamma  |E|^2\right)$, where $\Gamma=\gamma P d$, $\gamma$ is the nonlinear coefficient of the optical waveguides, $P$ is the peak input power and $E$ is the normalized field amplitude.  The evolution equations so obtained are a generalization of the one-ring Ikeda model \cite{ikeda} to a chain of coupled resonators: 
\begin{eqnarray}
A_{n}^{k+1}&=\left[j\theta A_{n-1}^k\exp\left(-j\Gamma\left|A_{n-1}^k\right|^2\right)+\right. \nonumber\\
&\left. \rho B_{n}^k\exp\left(-j\Gamma\left|B_{n}^k\right|^2\right)\right]\exp\left(-j\Omega\right)\exp\left(-j\Gamma\left|A_n^{k+1}\right|^2\right)\nonumber\\
B_{n}^{k+1}&=\left[j\theta B_{n+1}^k\exp\left(-j\Gamma\left|B_{n+1}^k\right|^2\right)+\right. \nonumber\\
&\left. \rho A_{n}^k\exp\left(-j\Gamma\left|A_{n}^k\right|^2\right)\right]\exp\left(-j\Omega\right)\exp\left(-j\Gamma\left|B_n^{k+1}\right|^2\right).\label{eq:modelo}
\end{eqnarray} 

Since the system response is frequency-periodic, $\Omega$ can be considered as a frequency detuning parameter from one reference resonance. For the analysis in terms of linear and nonlinear Bloch modes, a continuous-wave input signal with frequency $\omega$ is assumed \cite{jopt} that has stationary solutions of the form    
\begin{equation}
\left(
\begin{matrix}
A_n^k\\
B_n^k
\end{matrix}
\right)=
\left(
\begin{matrix}
A\\
B
\end{matrix}
\right)\exp\left(jnQ\right).
\end{equation}

When this ansatz is plugged into \eqref{eq:modelo}, the nonlinear dispersion relation obtained is
\begin{equation}
|\theta|\cos\left[Q-\Gamma\left|A\right|^2\left(|f|^2-1\right)\right]=\sin\left[\Omega +\Gamma\left|A\right|^2\left(|f|^2+1\right)\right] \label{eq::disp0}
\end{equation}
with 
\begin{equation}
f=\dfrac{B}{A}.
\end{equation}

The stationary evanescent solutions within the bandgaps can be shown to fulfill the condition $f=1$ \cite{jopt}. 

In the linear case, $\Gamma=0$ the well known linear dispersion relation
\begin{equation}
|\theta|\cos\left[Q\right]=\sin\left[\Omega\right] \label{eq::dispOL}
\end{equation}
is recovered.

A particularly interesting regime is found in the tight-binding limit $|\theta|\to1$ \cite{jopt}.   It has been shown that, for certain parameter values, a oscillatory response is obtained within the bandgap that corresponds to the self-structuring of the microring CROW field into spontaneously generated soliton trains.  

Even in the tight-binding regime, a CW input signal typically leads to an eventual irregular or chaotic regime after long evolution times.  This is something expected, since energy is continuously injected in a nonlinear structure where it recirculates for long times.  Nevertheless,  if an isolated pulse similar to the spontaneously generated solitons is re-injected in the transmission line, it propagates stably over very long distances \cite{jopt}.

The numerical scheme used for solving the time-domain evolution has been tested in the solution of a number of related cases where linear or nonlinear propagation takes place in microring CROWs \cite{jopt,ol2009,fraile}. In the evolution algorithm, each ring waveguide is split in a number of smaller sections and, therefore, the propagation of signals with a larger bandwidth than that restricted by the model equations \eqref{eq:modelo} is allowed for.    The effect of the reflection at the boundaries in finite structures was studied in detail in \cite{jopt}.  In this work, semi-infinite chains free from this effect will always be considered.

\section{Periodic gain and loss:  Type I structure}

\begin{figure}
\centerline{\includegraphics[width=.9\columnwidth]{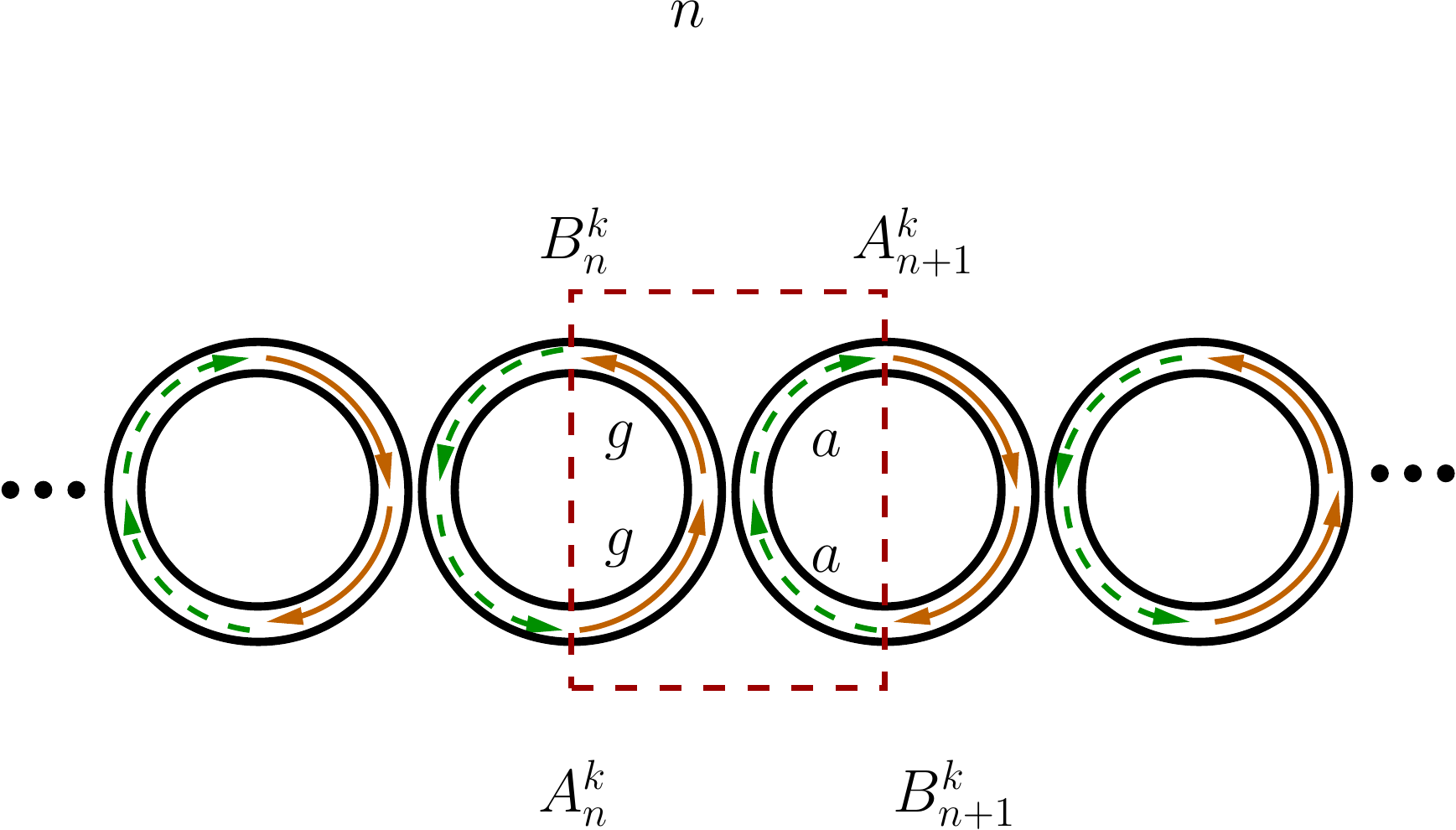}}
\caption{Type I structure with periodic gain and loss} \label{fig::crowI}
\end{figure}

The effect of gain and loss in the geometry displayed in Figure \ref{fig::crowI} is now considered.  In this CROW, each ring is divided into two sections and a periodic alternation of sections with gain $g$ (solid arrows) and loss $a$ (dashed arrows) is introduced.  The existence of a perfect balance $g=1/a$ between gain and loss will always be assumed.   The gain/loss periodicity in this case coincides with that of the microring CROW structure.   A parity inversion $P$ with center at any of the defined unit cell will have the effect of swaping the ordering of the gain and loss sections, while a time reversal $T$ will have the effect of turning gain into loss and loss into gain.  Therefore, the geometry depicted in Figure \ref{fig::crowI} is parity-time symmetric.

The perfect cancellation of propagation gain and loss within each ring, and the existence of additional losses due to the couplers, results in unconditionally stable transmission.  The possibility of having $a>1$ (and, therefore, $g<1$) which would correspond to the situation obtained by interchanging the gain and loss sections will also be considered.

An approximate exponential variation for the loss and amplification effects, $a=\exp(-\alpha d)$ and $g=\exp(\alpha d)$, will be assumed. The presence of gain or loss produces a modification of the nonlinear effects that is conventionally accounted for through the introduction of an effective length for each section.  The corrected nonlinear phase shifts in each case are  
\begin{equation}
\Gamma_a=\gamma P_0 d \dfrac{a^2-1}{2 \log a} \label{gammaa}
\end{equation}
and
\begin{equation}
\Gamma_g=\gamma P_0 d \dfrac{g^2-1}{2 \log g},\label{gammab}
\end{equation}
that correspond to an effective increase of the nonlinear effect in the gain sections and a decrease of the effective nonlinearity in the loss sections.

\subsection{Model Equations}

When the effects of gain and loss on the amplitudes of the propagating signals and the corrected effective nonlinearity coefficients \eqref{gammaa} and \eqref{gammab} are introduced.  The spatio-temporal evolution of the signal in the CROW is then described by the equations
\begin{eqnarray}
A_n^{k+1}&=&a\left[j\theta g A_{n-1}^{k}\exp\left(-j\Gamma_g\left|A_{n-1}^k\right|^2\right)\right.\nonumber\\
&&\left.+\rho a B_{n}^k\exp\left(-j\Gamma_a\left|B_n^k\right|^2\right)\right]\nonumber\\
&&\exp\left(-j\Omega\right)\exp\left(-j\Gamma_a\left|A_n^{k+1}\right|^2/a^2\right)  \nonumber\\
B_n^{k+1}&=&g\left[j\theta a B_{n+1}^{k}\exp\left(-j\Gamma_a\left|B_{n+1}^k\right|^2\right)\right.\nonumber\\
&&\left.+\rho g A_{n}^k\left(-j\Gamma_g\left|A_n^k\right|^2\right)\right]\nonumber\\
&&\exp\left(-j\Omega\right)\exp\left(-j\Gamma_g\left|B_n^{k+1}\right|^2/g^2\right) \label{eq::modelI}
\end{eqnarray}

\subsection{Band structure}

We now assume a stationary solution of the type
\begin{equation}
\left(
\begin{matrix}
A_n^k\\
B_n^k
\end{matrix}
\right)=
\left(
\begin{matrix}
A\\
B
\end{matrix}
\right)\exp\left(jnQ\right).
\end{equation}

When the above ansatz is introduced in the model equation \eqref{eq::modelI}, the nonlinear dispersion relation in this structure obtained is
\begin{equation}
|\theta|\cos\left[Q-\Gamma_g\left|A\right|^2\left(|f_a|^2-1\right)\right]=\sin\left[\Omega +\Gamma_g\left|A\right|^2\left(|f_a|^2+1\right)\right] \label{eq::dispI}
\end{equation}
with 
\begin{equation}
f_a=a\dfrac{B}{A}.
\end{equation}

\begin{figure}
\centering
\begin{tabular}{cc}
(a)&(b)\\
\includegraphics[width=.45\columnwidth]{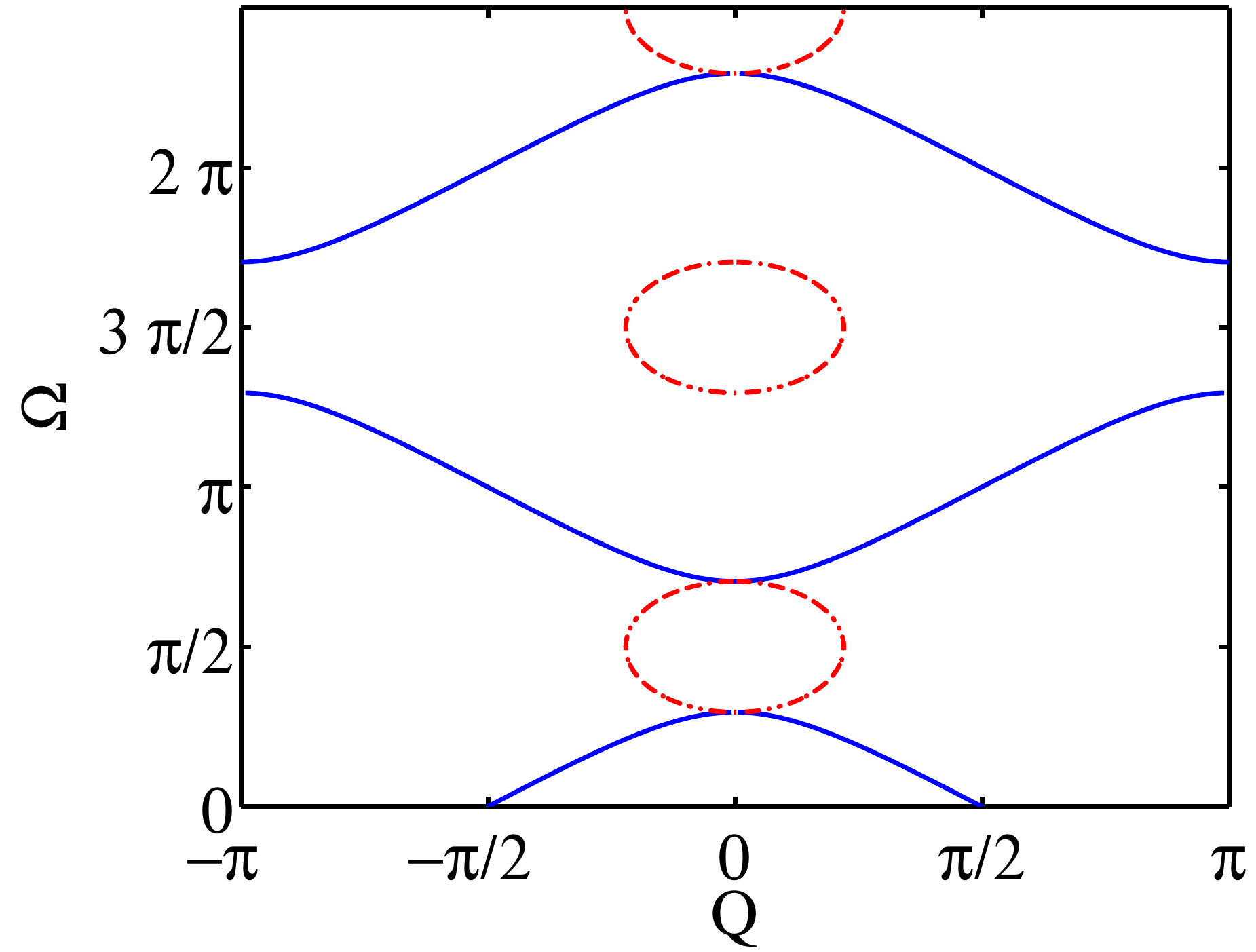}&\includegraphics[width=.45\columnwidth]{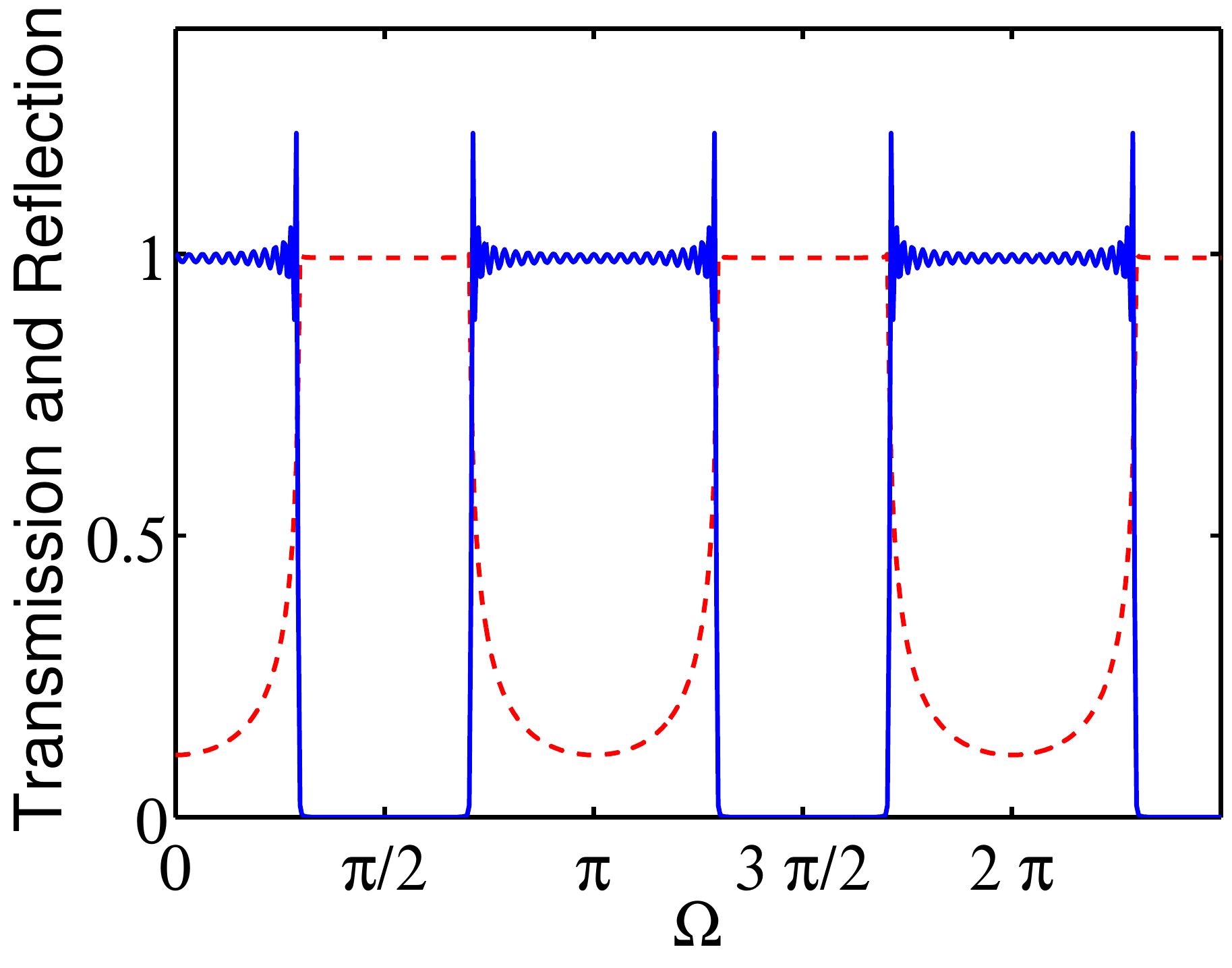}
\end{tabular}
\caption{(a) Band structure for a type I linear microring CROW with periodic gain and loss, $|\theta|=0.8$ and $g=10$ and (b) the transmission (solid line) and reflection (dashed-line) spectra for an input impulse.}\label{fig:dispersionI}
\end{figure}

The corresponding expression for the linear case is obtained by setting $\Gamma_g=0$ in Equation \ref{eq::dispI} and is identical to that of the structure without gain or loss. Figure \ref{fig:dispersionI} (a) shows the band structure for linear a type I microring CROW with periodic gain and loss and  $|\theta|=0.8$.  Figure \ref{fig:dispersionI} (b) displays the normalized spectra of the transmitted (solid line) and reflected (dashed line) signal obtained form an input unit impulse. The computed results show clearly the boundaries of the forbidden propagation bands.  The ringing observed in the spectra are due to the truncation of the unit response due to the finite length of the computer simulation and the long lived nature of the oscillations in the structure.  

\subsection{Propagation in type I structures}

\begin{figure}
\centering
\begin{tabular}{cc}
(a)&(b)\\
\includegraphics[width=.45\columnwidth]{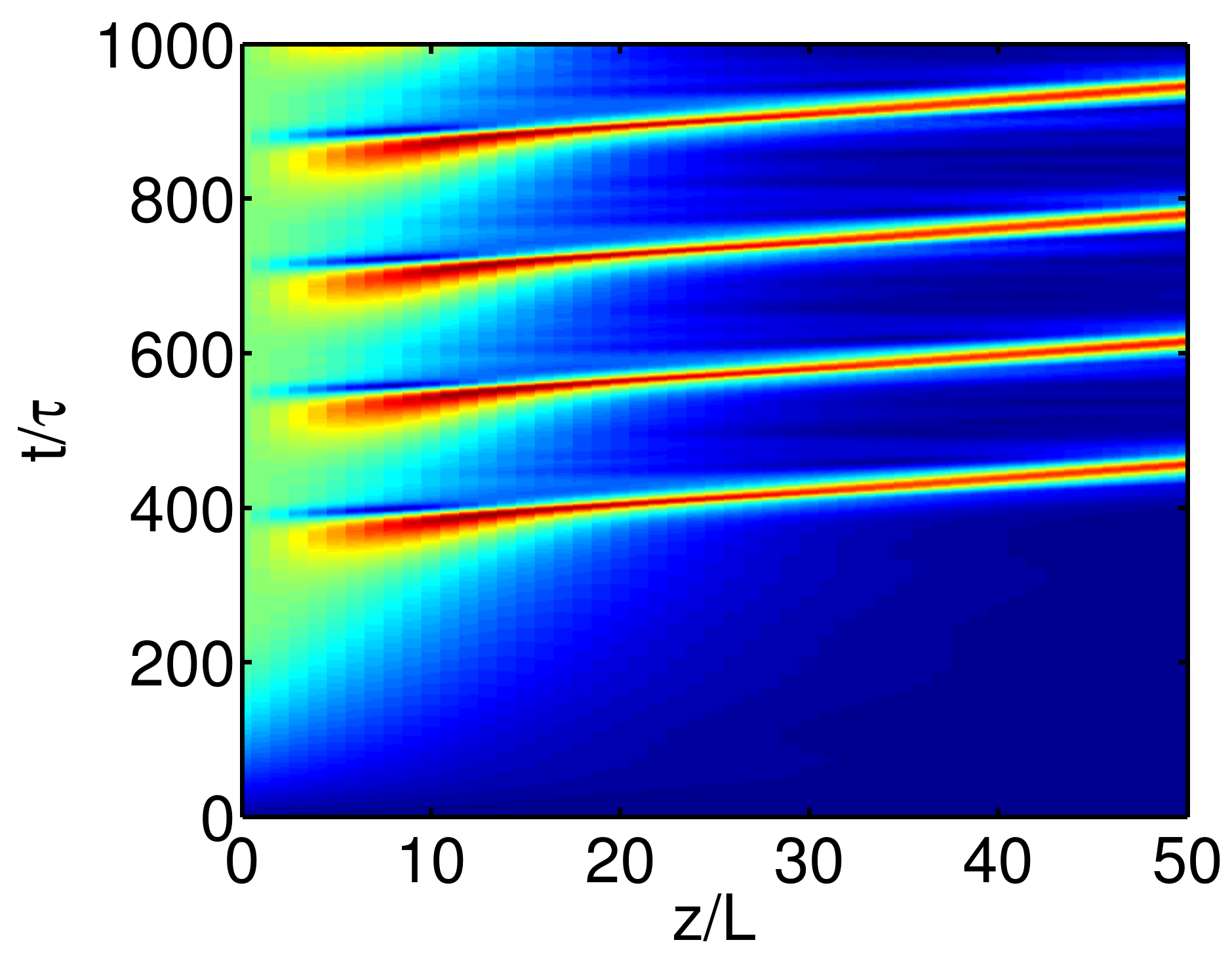}&\includegraphics[width=.45\columnwidth]{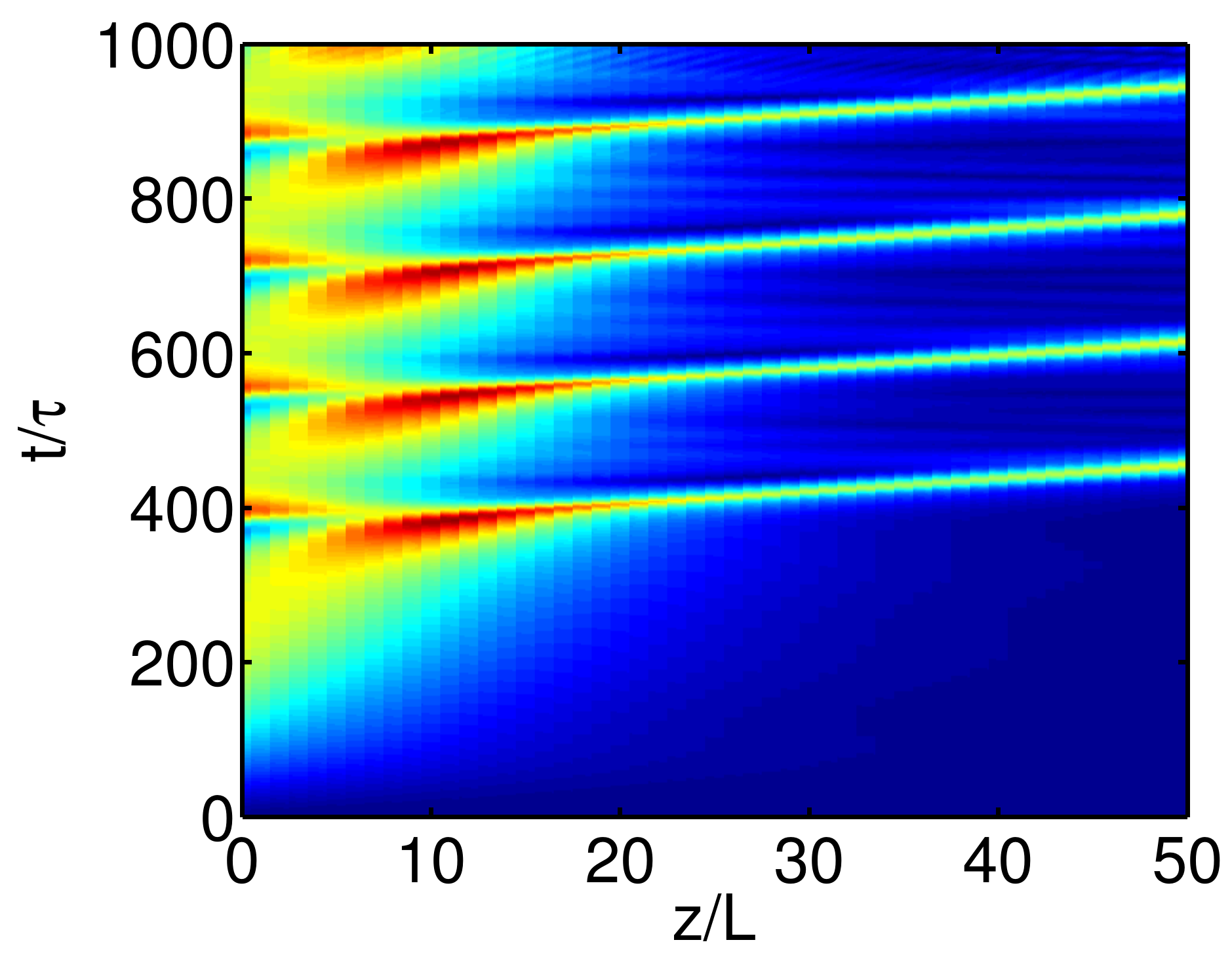}\\
(c)&(d)\\
\includegraphics[width=.45\columnwidth]{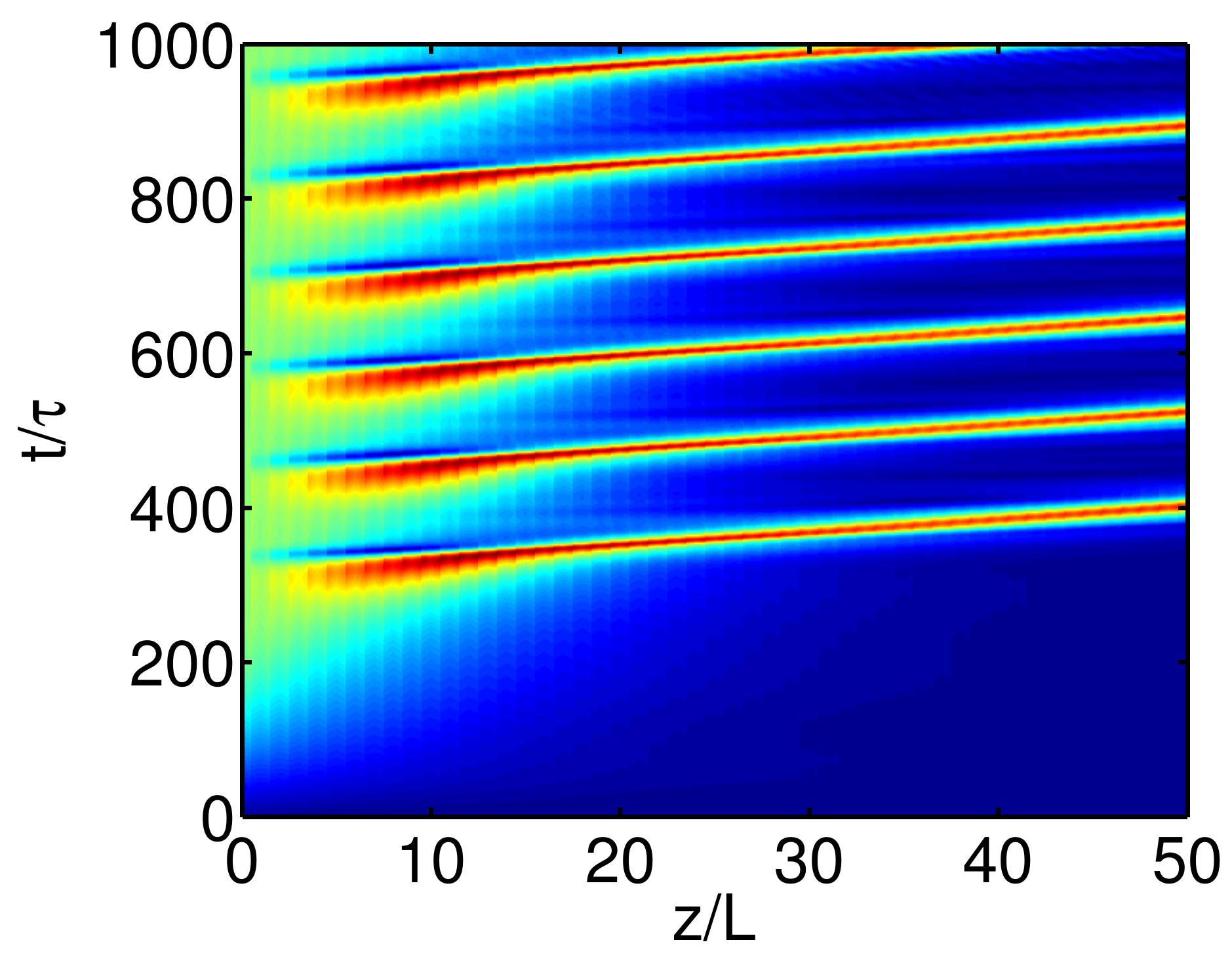}&\includegraphics[width=.45\columnwidth]{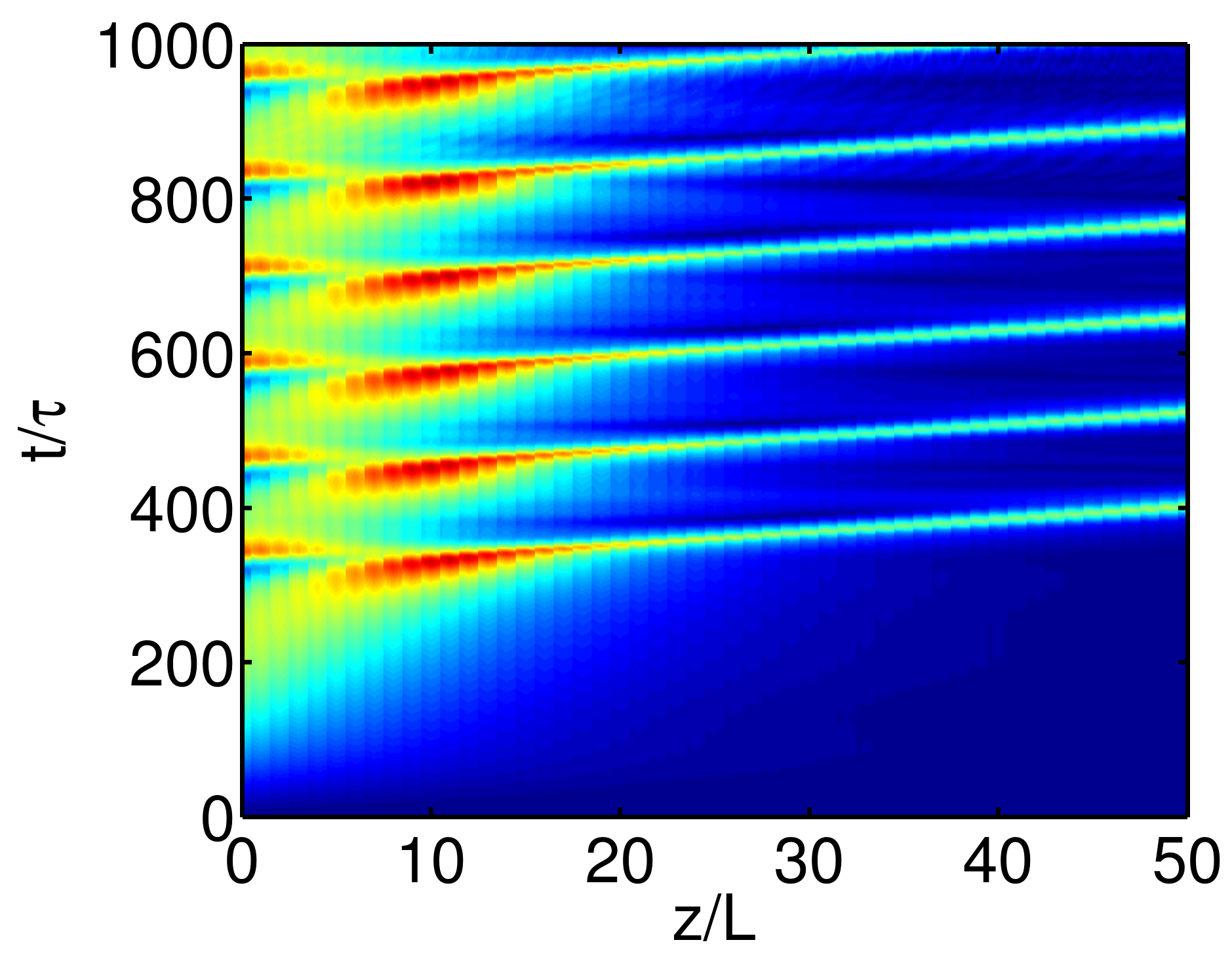}\\
(e)&(f)\\
\includegraphics[width=.45\columnwidth]{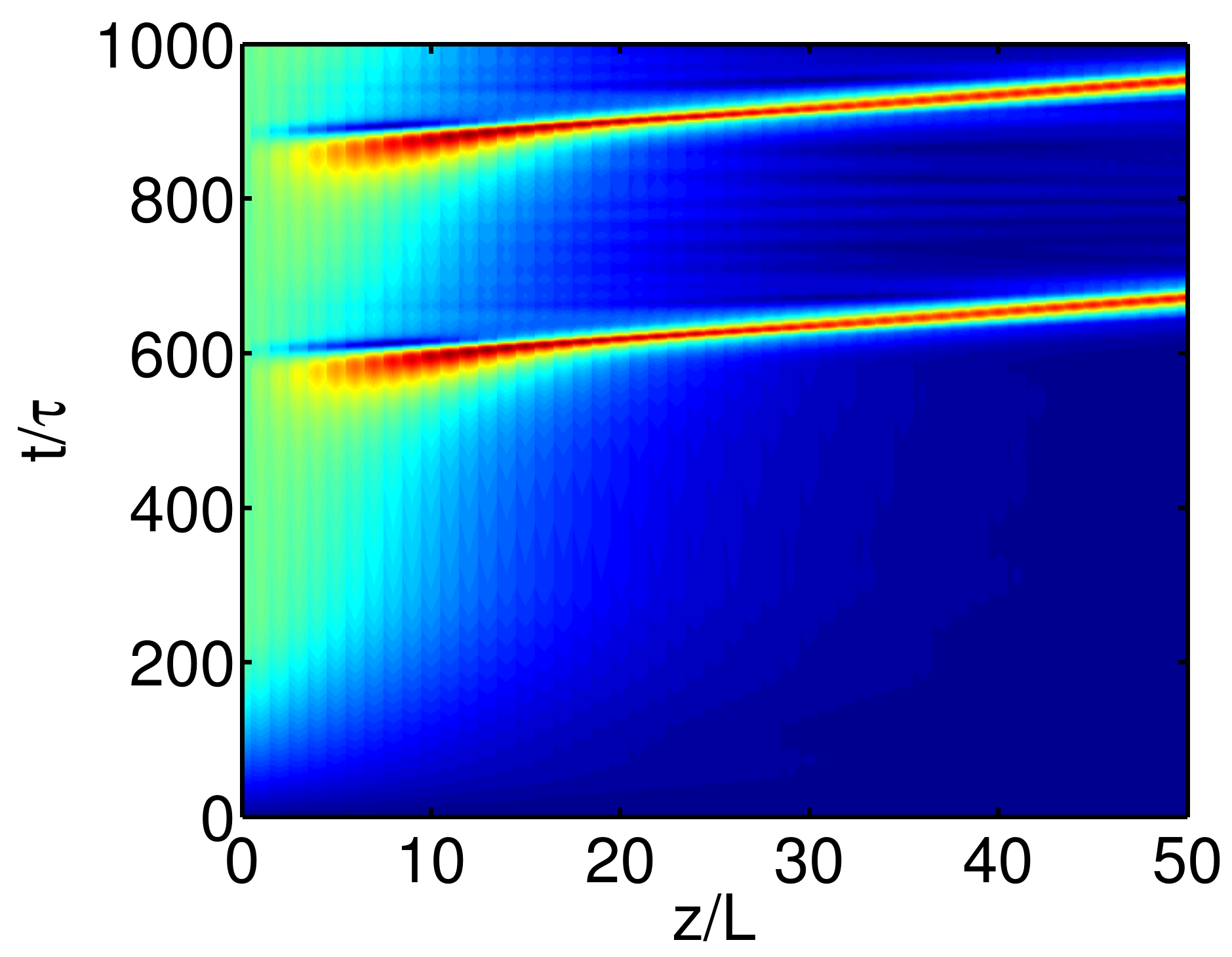}&\includegraphics[width=.45\columnwidth]{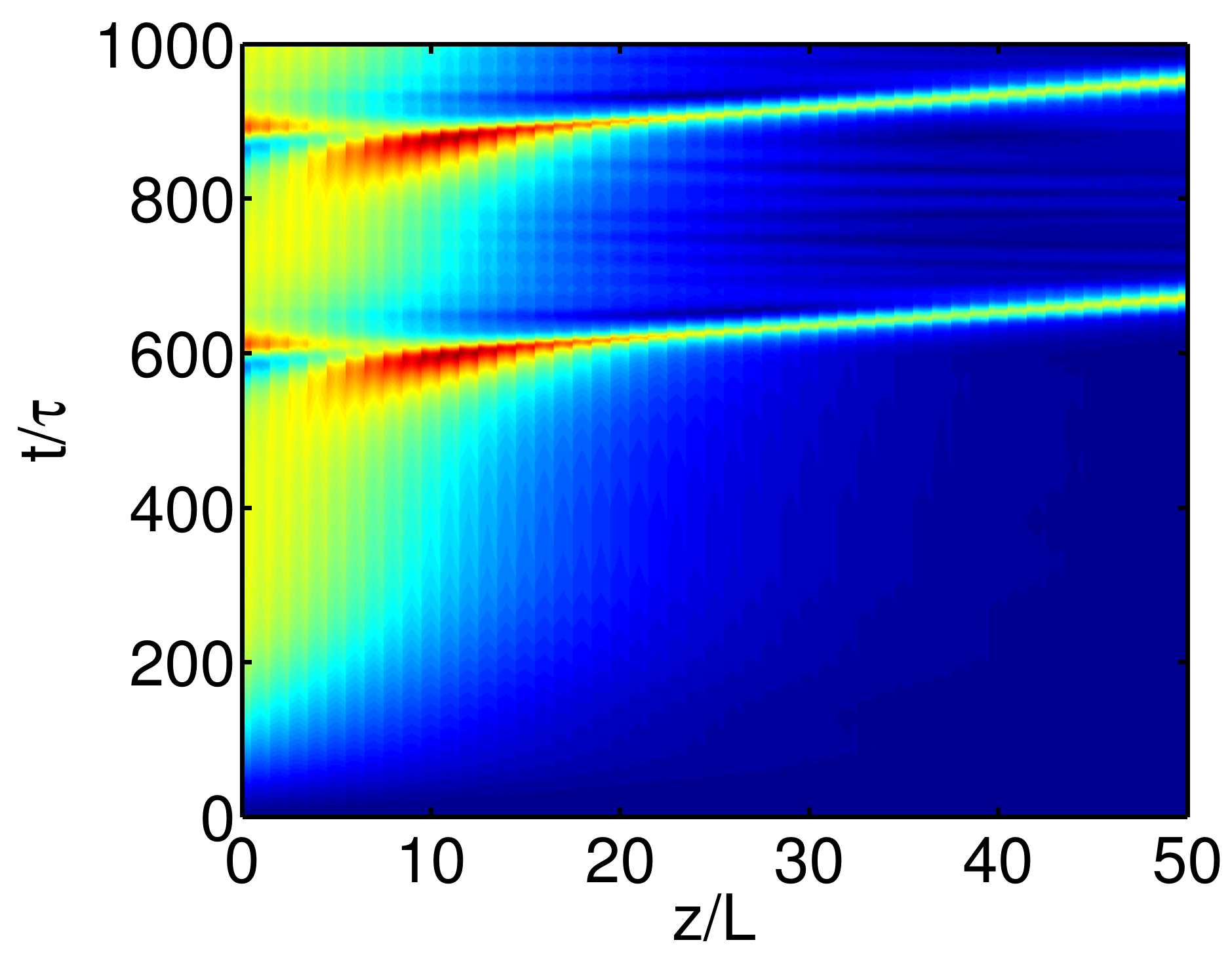}
\end{tabular}
\caption{Spontaneous generation of soliton trains from an input CW step signal when $|\theta|=0.992$ and $\Gamma=0.055$.  The $|A|$ component is shown in (a) for $g=1/a=1$, (c) for $g=1/a=1.1$ and (e) for $g=1/a=0.9$.  The corresponding cases for the $|B|$ component are shown in (b), (d) and (f)}\label{fig::Geff}
\end{figure}

For each microresonator, the total loop amplification factor is $a^2g^2(1-\theta^2)$. The gain/loss balance $ag=1$ together with the condition $(1-\theta^2)\le1$ leads to 
\begin{equation}
a^2g^2(1-\theta^2)\le 1,
\end{equation}
which ensures the stability against a lasing transition at any individual microring.  This is a necessary condition for the stability of the solutions propagating in the structure.

The linear dispersion relations of type I structures are identical to those of the microring CROW without gain or loss.  In particular, no symmetry breaking transition is observed as the gain and loss is varied.  In the nonlinear case,  there are two net effects arising from the presence of periodic gain and loss that can be observed from the direct comparison of Equations \eqref{eq::disp0} and \eqref{eq::dispI}.  The first one is a modification of the effective nonlinearity dictated by the transformation $\Gamma\to \Gamma_g$ in \eqref{eq::dispI}.   The second effect is the introduction of a local imbalance between the forward and backward propagating field components of the stationary nonlinear Bloch modes.

The effective change of the nonlinearity with the gain/loss coefficient is illustrated in Figure \ref{fig::Geff} where the amplitude of the propagating $A$ and $B$ components obtained from a CW input with  $|\theta|=0.992$ and $\Gamma=0.055$ are shown for three different values of the gain/loss parameter $g=1/a$. A CW input step with a smooth leading edge tuned to a ten percent of the total frequency gap inside the band gap has been used.  The carrier frequency is the same for all the computations in this work. Figure \ref{fig::Geff} (a) and (b) show the results for $g=1$.  When the gain/loss parameter is increased to $g=1/a=1.1$ we observe in   Figures \ref{fig::Geff} (c) and (d) a increase in the repetition rate of the output pulses, which typically corresponds to an increase of the nonlinearity parameter in the absence of gain and loss \cite{jopt}.   

Similarly to the the case of the structures without gain or loss, we have \cite{jopt} that for the evanescent nonpropagating solutions within the band gap $|f_a|=1$.  The condition $|f|=1$ when $a=g=1$ results in equal amplitudes for the $A$ (forward) and $B$ (backward) components of the nonpropagating Bloch modes.  When $a<1$, the condition $|f_a|=1$ implies that $|B|>|A|$.  This permits to compensate the variations of the optical field along the structure an to have, on average, equal contributions from the forward and backward components.  Also,the ratio of the peak values of $A$ and $B$ observed in the spontaneously generated soliton trains is the same irrespective of the value of the parameter $g=1/a$, even though its value affects the repetition rate of the generated solitons.

The increase of the effective nonlinearity with $g$ is due to the fact that in the definition of the type I structure in Figure \ref{fig::crowI} gain has been considered in the first half of the unit cell.  Therefore, the first section faced by the optical field when it propagates in each cell of the periodic structure is a gain section and this enhances the effective nonlinearity.  The converse situation can be considered simply by setting $g<1$.  In this case the first part of the cell period corresponds to the propagation in the presence of loss and there is a reduction of the effective nolinearity.  This is illustrated in Figures \ref{fig::Geff} (e) and (f) , which show the pulse rate decrease typical of a smaller nolinearity. Similarly, the accompanying condition $a=1/g>1$ and $|f_a|=1$ implies that $|A|>|B|$ in the band-gap solutions if the ordering of the gain and loss section is reversed.

\section{Periodic gain and loss: Type II structures}

\begin{figure}
\centering
\centerline{\includegraphics[width=.9\columnwidth]{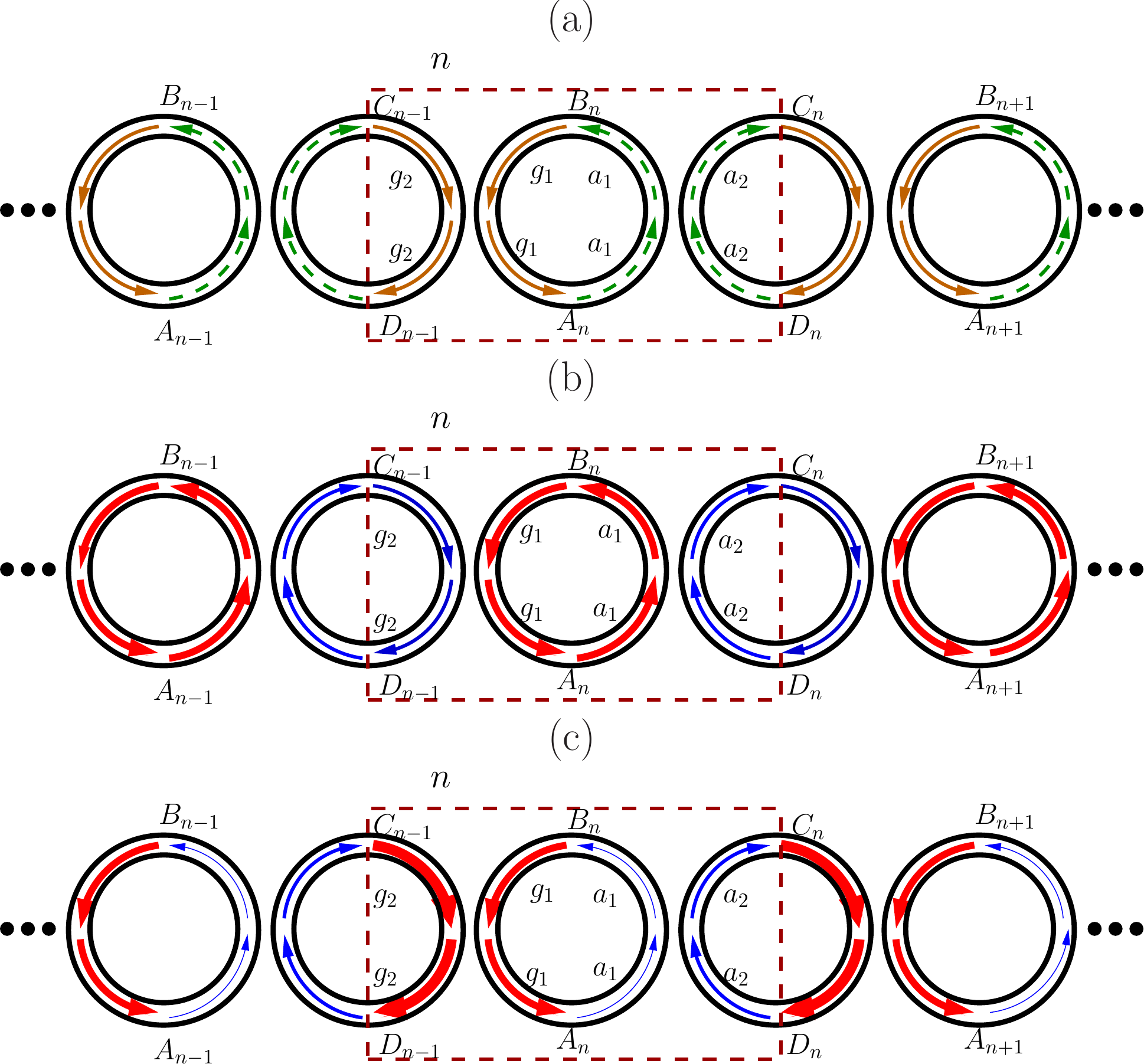}}
\caption{(a) Generic Type II structure with periodic gain and loss. (b) An example with alternating full active and lossy microrings. (c) An example with each microring split in two parts with gain and loss. }\label{fig::crowII}
\end{figure}

Another structure with the geometry shown in Figure \ref{fig::crowII} (a) is considered.  Solid and dashed arrows correspond to gain and loss sections, respectively.  In this setup there is a doubling of the period of the gain and loss variation in relation with the fundamental period of the microring sequence.  Again, it will be assumed that there is a perfect gain-loss balance in a full period for the gain-loss variation, which leads to the restriction
\begin{equation}
a_1g_1a_2g_2=1. \label{eq::restriction}
\end{equation}
A perfect gain/loss balance will not be assumed within each individual microring, but restriction \eqref{eq::restriction} imposes the gain/loss balance within a two-ring cell that constitute the gain/loss period.  This permits to define a gain/loss parameter
\begin{equation}
k  \equiv \left(a_2 g_2\right)^2=\dfrac{1}{\left(a_1 g_1\right)^2}. \label{eq::glparam}
\end{equation}

Also, the analysis will again be assumed to be valid for each individual $a_l$ or $g_l$ greater or smaller than $1$, so each individual section in Figure \ref{fig::crowII} (a) can actually correspond to the presence of either gain or loss in that specific transmission section, but always within the limits of the restrictions of \eqref{eq::glparam} which also implies the fulfillment of \eqref{eq::restriction}.  

In structures of the type II we do observe symmetry breaking transitions.  In fact, the interchangeability assumed for the gain and loss sections results in the existence of two such transitions, which are symmetrical in the gain-loss parameter $k$.

There are several possible orderings of the gain and loss sections that fall within the definition of type II structures.  Figure \ref{fig::crowII} (b) corresponds to one such ordering in which $a_1=g_1$ and $a_2=g_2$.  In this case, we have 

\begin{equation}
k=g_2^4=1/a_1^4,
\end{equation}
which can be either greater or smaller than one depending on the specific ordering gain-loss-gain-loss... or loss-gain-loss-gain..., respectively.  In either case, there is an alternation of microrings which are either active or lossy along their full length.  For an infinite length chain, we take a parity $P$ reflection relative to a reference point placed at the coupling between two microrings.  This transformation, when combined with a time reversal $T$ that exchanges gain and loss sections, leaves the structure invariant.  Therefore, we have a parity-time symmetry in this structure that can be considered as an arbitrary order extension of the two ring systems studied in \cite{peng} and \cite{bender2013}. 

Another possibility is shown in Figure \ref{fig::crowII} (c).  Here, the amplitude of the gain or loss section is indicated by the thickness of the corresponding arrow and there is an alternation of gain and loss sections within each ring.  In this case we have $a_2=1/g_1$ and $a_1=1/g_2$, which permits to fulfill the condition \eqref{eq::restriction} for the gain-loss balancing across two consecutive microrings.  In this case, the definition of the gain/loss parameter \eqref{eq::glparam} leads to 
\begin{equation}
k=\left(\dfrac{g_2}{g_1}\right)^2=\left(\dfrac{a_2}{a_1}\right)^2.
\end{equation}
Again, since we admit $g_l$ and $a_l$ both greater or larger than $1$, $k$ will take values smaller and greater than $1$ and two symmetry-breaking transitions will be found.

\subsection{Model equations}

To accommodate all the internal degrees of freedom of the optical field within one period of the gain/loss variations, we need four field variables.  Two for the forward propagating components ($A$ and $C$) and two for the backward propagating components ($B$ and $D$).   The evolution equations for type II structures are 
\begin{eqnarray}
A_n^{k+1}&=&g_1\left[j\theta g_2 C_{n-1}^{k}\exp\left(-j\Gamma_{g_2}\left|C_{n-1}^k\right|^2\right)+\right.\nonumber\\
&&\left.\rho g_1 B_{n}^k\exp\left(-j\Gamma_{g_1}\left|B_n^k\right|^2\right)\right]\nonumber\\
&&\exp\left(-j\Omega\right)\exp\left(-j\Gamma_{g_1}\left|A_n^{k+1}\right|^2/g_1^2\right)\nonumber\\
B_n^{k+1}&=&a_1\left[j\theta a_2 D_{n}^{k}\exp\left(-j\Gamma_{a_2}\left|D_{n}^k\right|^2\right)+\right.\nonumber\\
&&\left.\rho a_1 A_{n}^k\exp\left(-j\Gamma_{a_1}\left|A_n^k\right|^2\right)\right]\nonumber\\
&&\exp\left(-j\Omega\right)\exp\left(-j\Gamma_{a_1}\left|B_n^{k+1}\right|^2/a_1^2\right)\nonumber\\
C_n^{k+1}&=&a_2\left[j\theta a_1 A_{n}^{k}\exp\left(-j\Gamma_{a_1}\left|A_{n}^k\right|^2\right)+\right.\nonumber\\
&&\left.\rho a_2 D_{n}^k\exp\left(-j\Gamma_{a_2}\left|D_n^k\right|^2\right)\right]\nonumber\\
&&\exp\left(-j\Omega\right)\exp\left(-j\Gamma_{a_2}\left|C_n^{k+1}\right|^2/a_2^2\right)\nonumber\\
D_n^{k+1}&=&g_2\left[j\theta g_1 B_{n+1}^{k}\exp\left(-j\Gamma_{g_1}\left|B_{n+1}^k\right|^2\right)+\right.\nonumber\\
&&\left. \rho g_2 C_{n}^k\left(-j\Gamma_{g_2}\left|C_n^k\right|^2\right)\right]\nonumber\\
&&\exp\left(-j\Omega\right)\exp\left(-j\Gamma_{g_2}\left|D_n^{k+1}\right|^2/g_2^2\right).\label{eq::modeloII}
\end{eqnarray}

When $a_2=a_1$ and $g_1=g_2$ this structure reduces to the previously described type I.  In this case, there is a correspondence of $C_n$ in the type II structure with $A_{n+1}$ in type I and, of $D_n$ with $B_{n+1}$ and, therefore, the apparent period doubling does not actually produce any new solution.

\subsection{Band Structure}

\begin{figure}
\centering
\begin{tabular}{cc}
(a)&(b)\\
\includegraphics[width=.45\columnwidth]{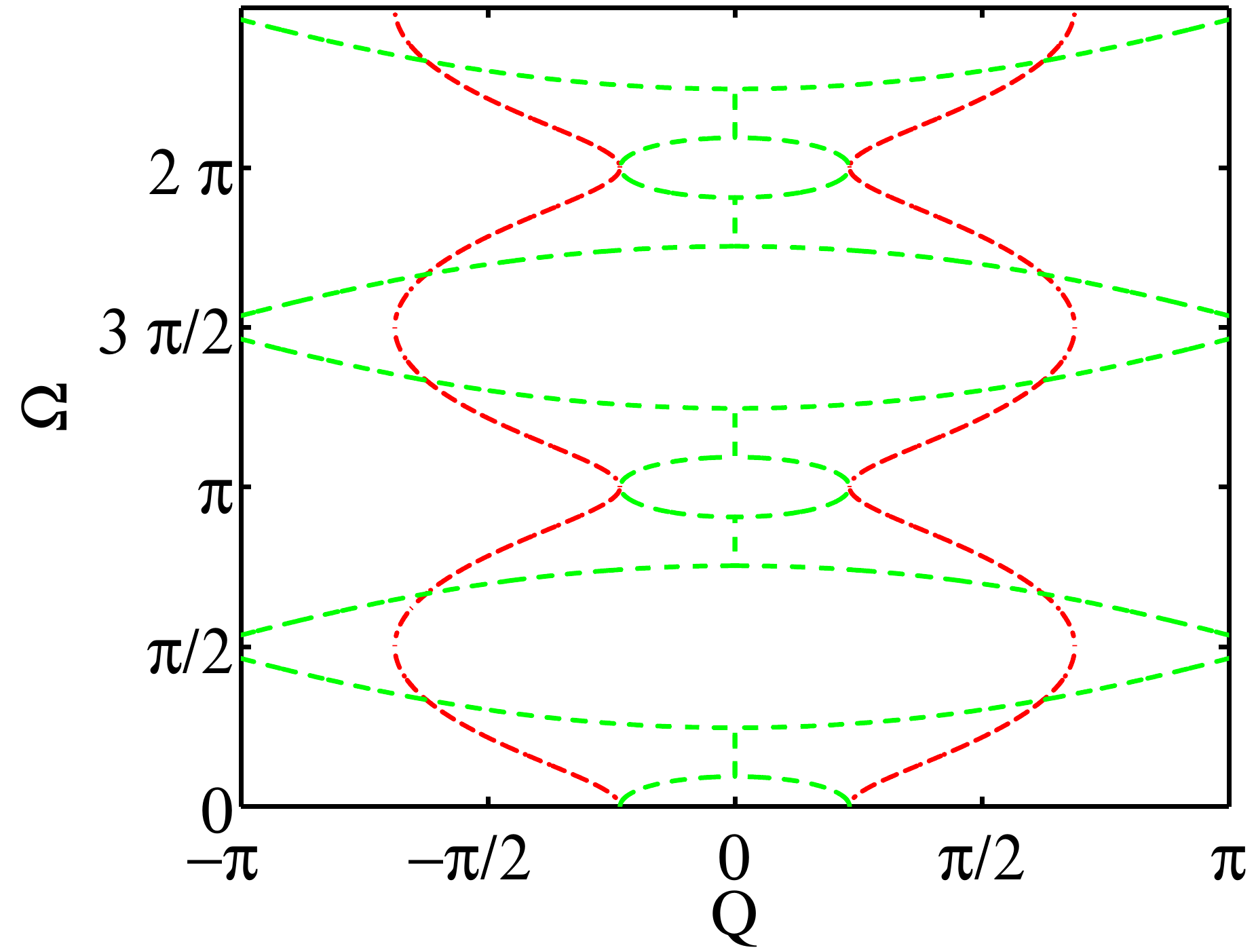}&\includegraphics[width=.45\columnwidth]{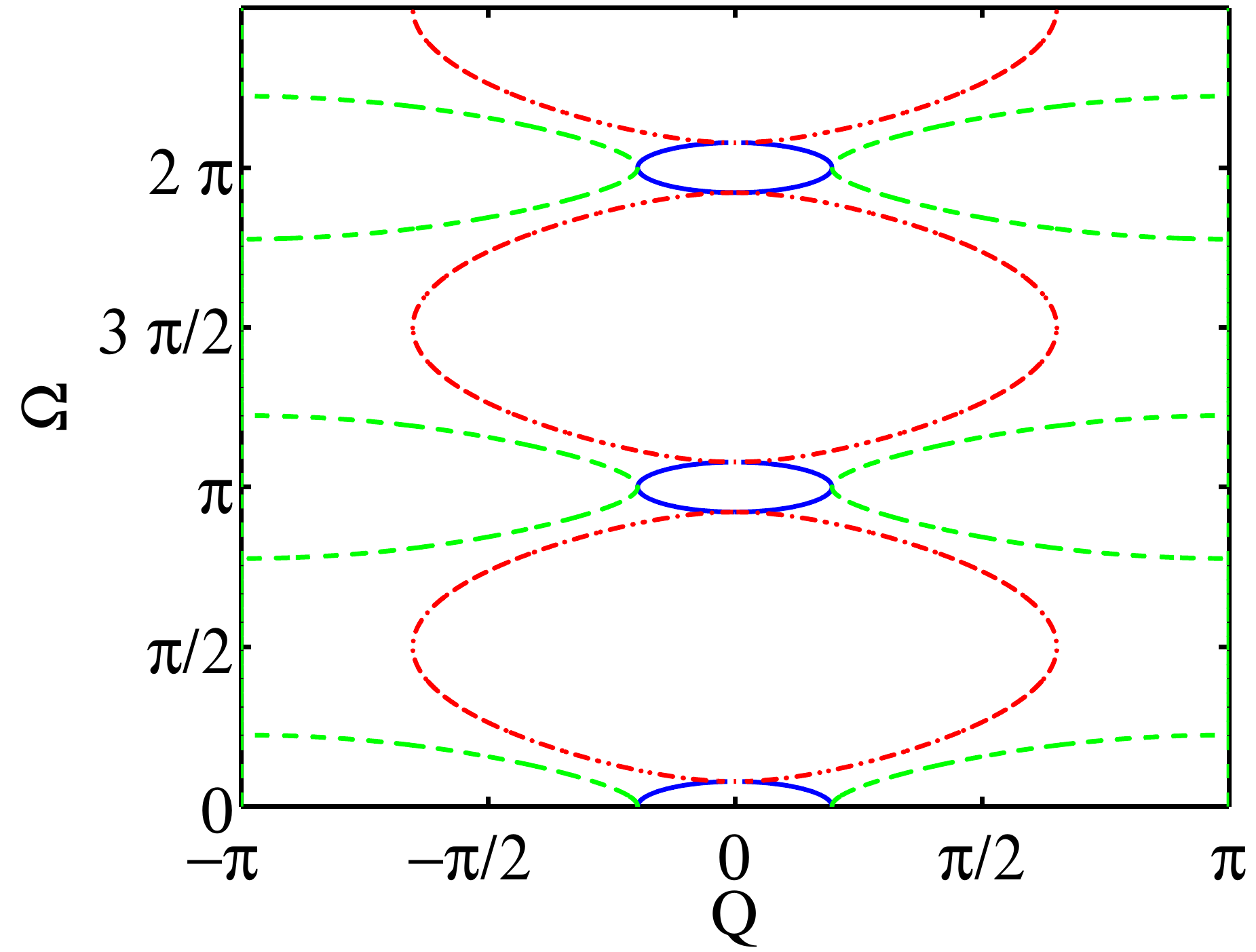}\\
(c)&(d)\\
\includegraphics[width=.45\columnwidth]{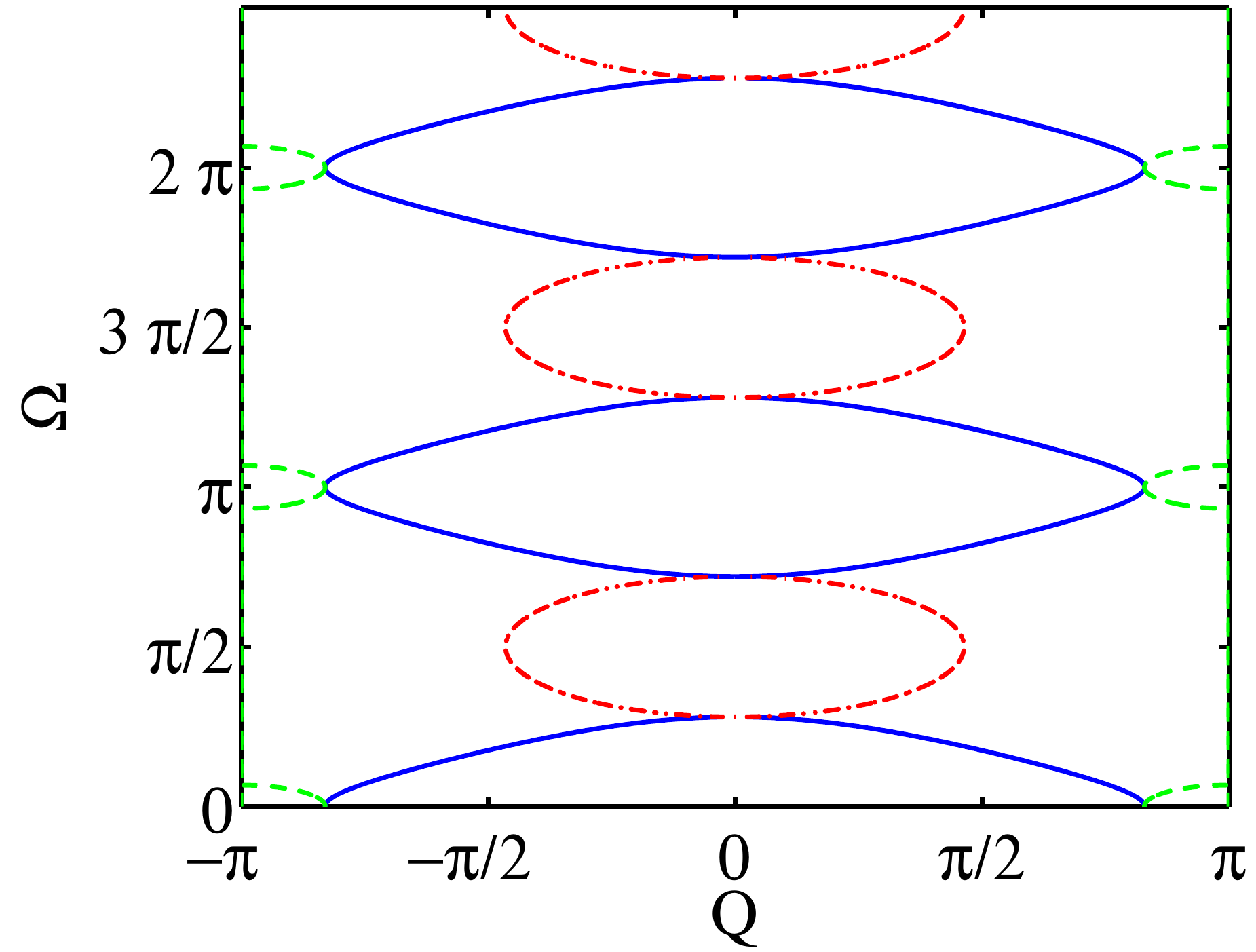}&\includegraphics[width=.45\columnwidth]{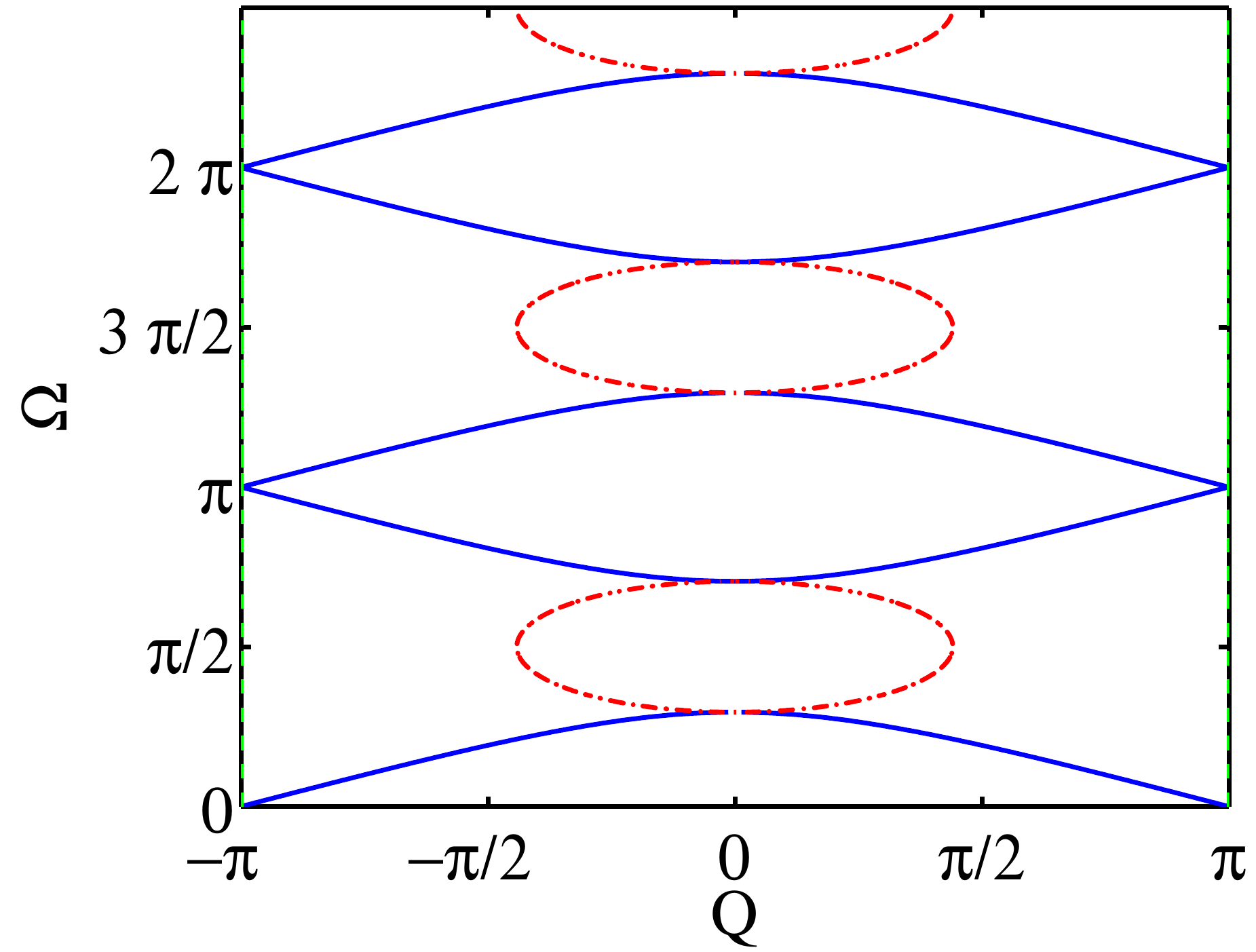}\\
(e)&(f)\\
\includegraphics[width=.45\columnwidth]{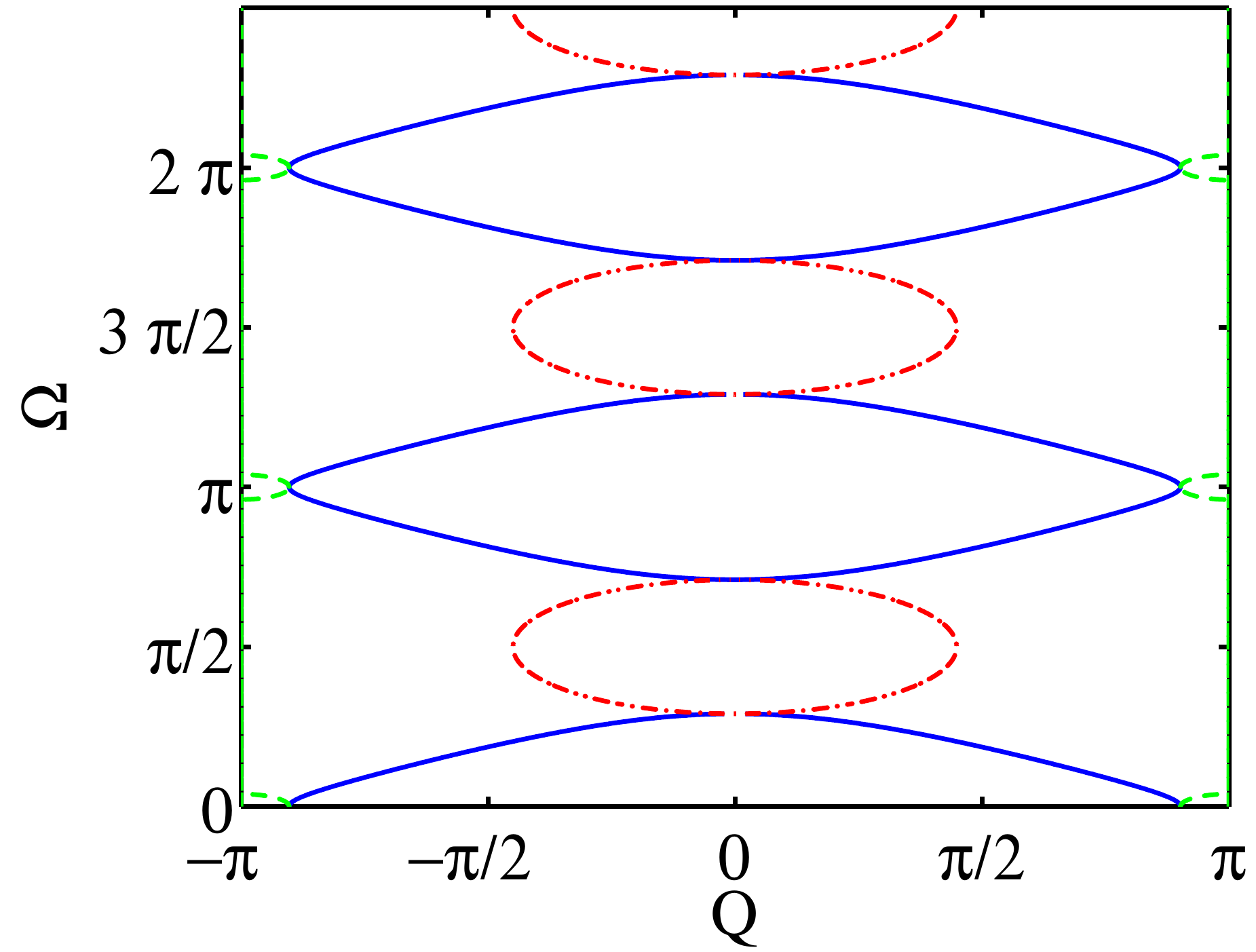}&\includegraphics[width=.45\columnwidth]{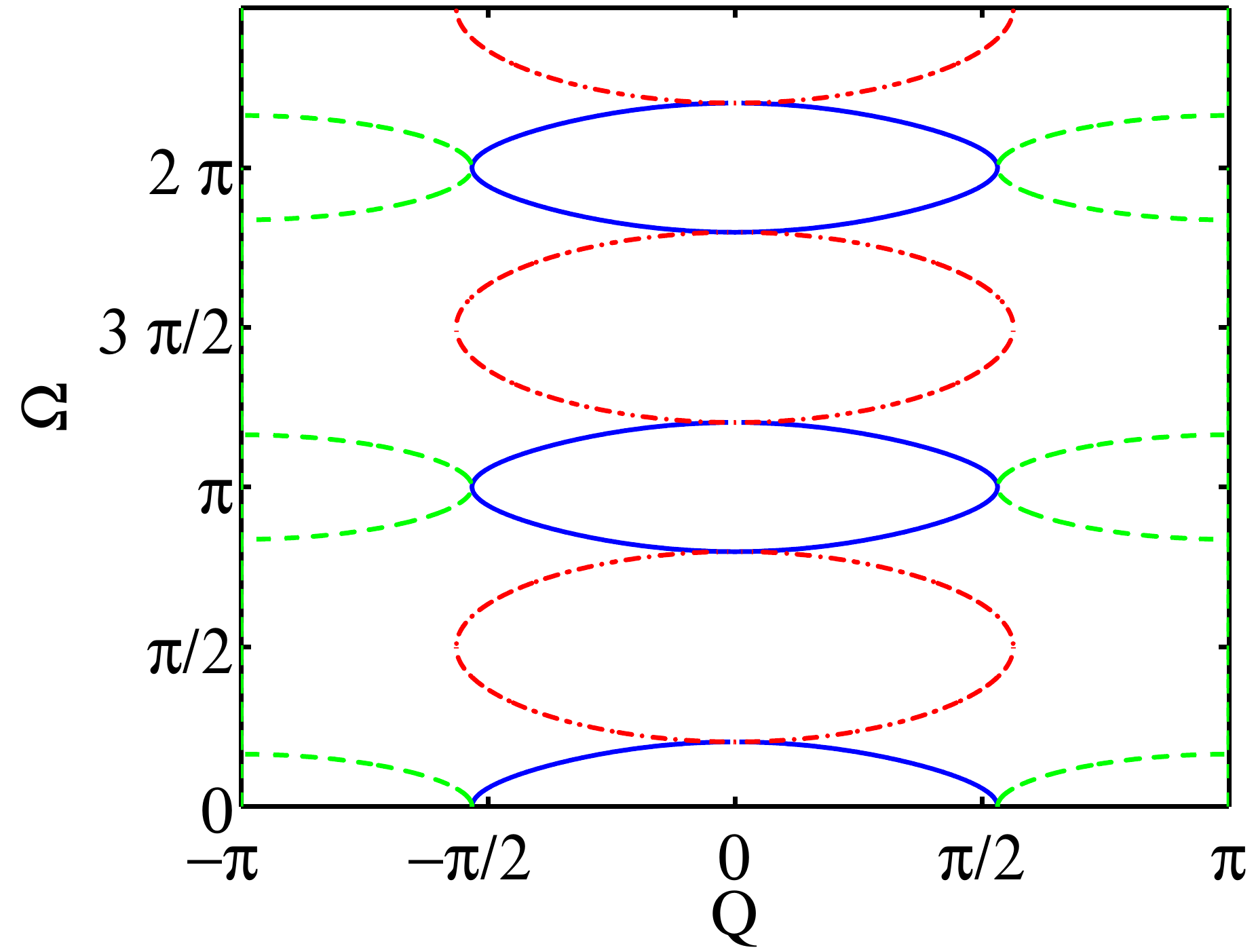}\\
(g)&(h)\\
\includegraphics[width=.45\columnwidth]{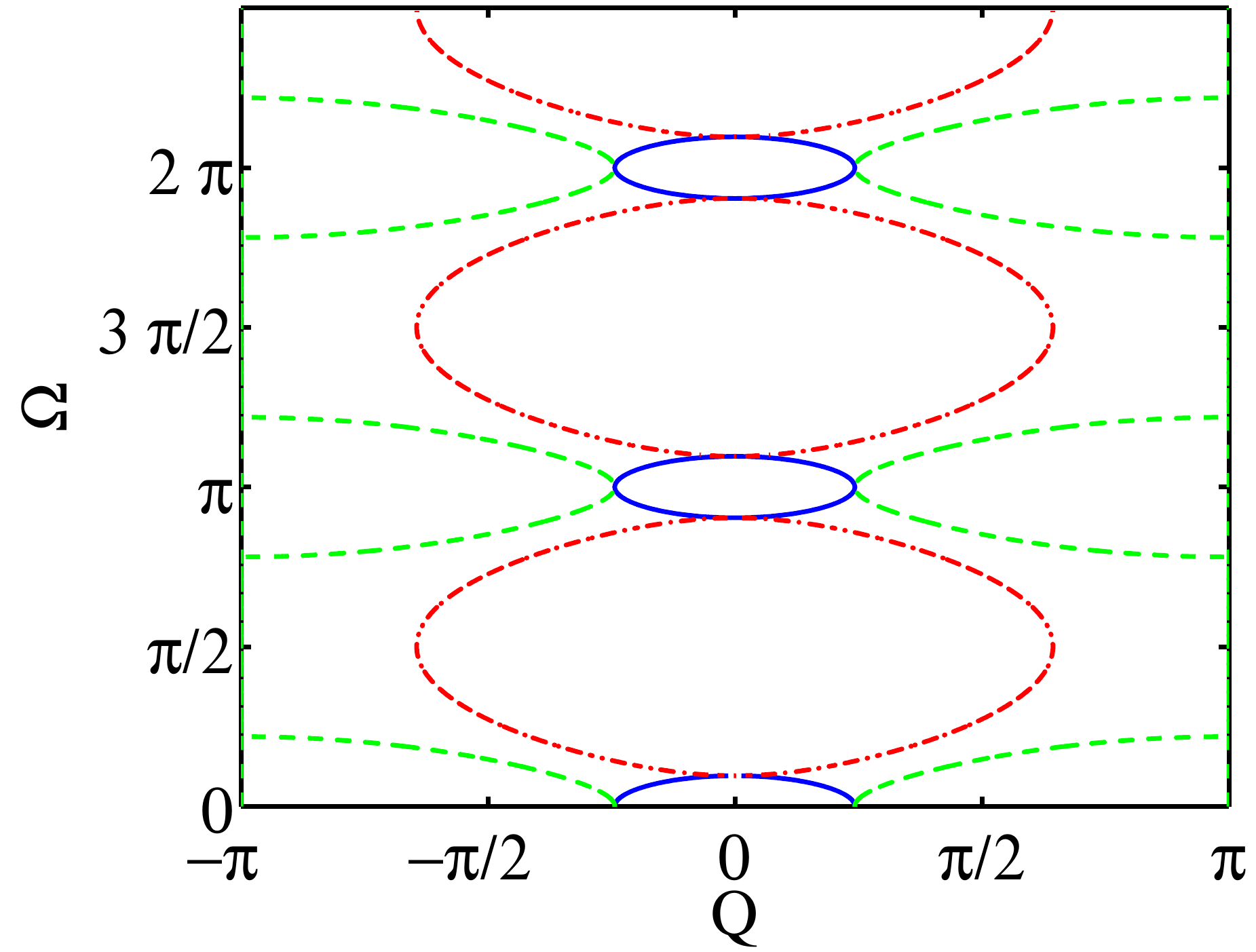}&\includegraphics[width=.45\columnwidth]{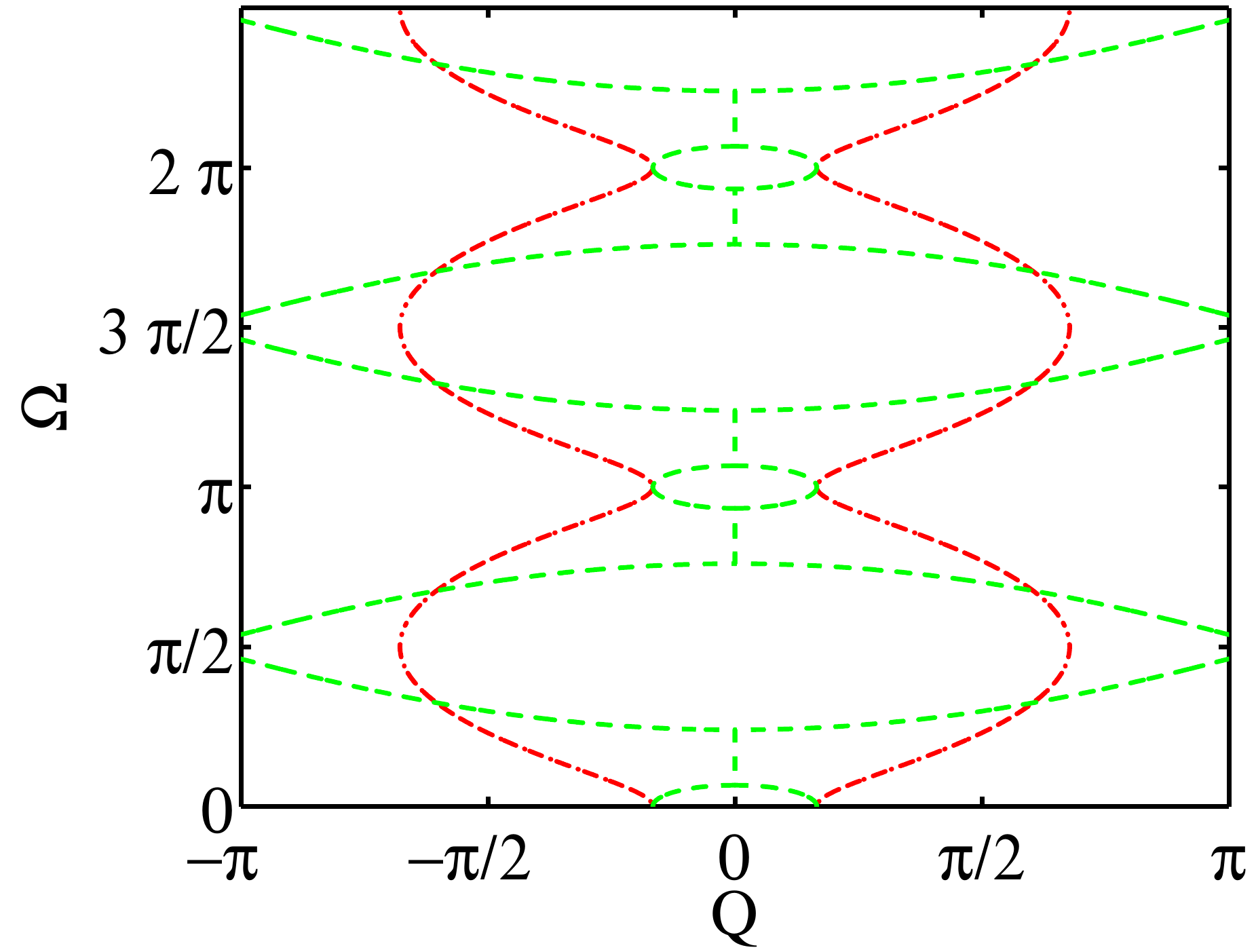}
\end{tabular}
\caption{Band structure for $|\theta|=0.8$ and eight different values of the gain/loss parameter $k$: (a) $k=0.1$, (b) $k=0.12$,  (c) $k=0.5$, (d) $k=1$, (e) $k=1.5$, (f) $k=5$,  (g) $k=8$, (h) $k=9.5$}\label{fig::ptbands}
\end{figure}

We consider now the Bloch-type stationary solutions as
\begin{equation}
\left(
\begin{matrix}
A_n^k\\
B_n^k\\
C_n^k\\
D_n^k
\end{matrix}
\right)=
\left(
\begin{matrix}
A\\
B\\
C\\
D
\end{matrix}
\right)\exp\left(jnQ\right).
\end{equation}

Introducing this ansatz in \eqref{eq::modeloII} we obtain the linear dispersion relation
\begin{equation}
\theta^2\cos\left(Q\right)=\dfrac{1-\theta^2}{2 k}\left(1+k^2\right)-\cos\left(2\Omega\right).\label{eq::disp}
\end{equation}

This dispersion relation depends only  the gain/loss parameter $k$ defined in \eqref{eq::glparam} and the coupling coefficient $\theta$.  It is noteworthy to point out that different organizations of the periodic loss-gain structure can lead to the same value of $k$ and, therefore, display identical fundamental propagation properties.  For instance, we obtain the same band structure if the gain-loss parameter has the same value for both the structures depicted in Figures \ref{fig::crowII} (b) and (c) even though the physical layouts are different. 

In the $k=1$ case, the dispersion relation of the type I structure is recovered, provided that the substitution $Q\to 2Q$ is made, since we are doubling the period of the structure in a trivial way.  Again, different physical implementations can lead to this particular limit case.  If the implementation depicted in Figure \ref{fig::crowII} (b) is used: alternation of fully active or lossy rings, $k=1$ corresponds to the absence of gain and loss and one recovers the setup of Figure \ref{fig::crow0}.  If we consider the specific arrangement in Figure \ref{fig::crowII} (c) having two gain and loss sections within a single ring, we obtain in the $k=1$ case an structure that is equivalent to the type I structure described above.  

A symmetry breaking transition is obtained when the propagation factor $Q$ in \eqref{eq::disp} becomes imaginary.  At $Q=0$, this defines the limits 
\begin{equation}
\dfrac{1-|\theta|}{1+|\theta|}\le k \le \dfrac{1+|\theta|}{1-|\theta|}.\label{eq::pt}
\end{equation}

The existence of two transitions instead of one at values of $k$ that are their respective inverses is a result of the fact that we considered the formal possibility that $a_l$ and $g_l$ correspond to either loss or gain, by having values either smaller or greater than $1$. 

The band structures for different values of $k$ and $|\theta|=0.8$ are shown in Figure \ref{fig::ptbands}.  The real $Q$ eigenvalues are shown with solid lines, whereas the imaginary eigenvalues are shown with dashed-dotted lines.  No real eigenvalues exist within the broken symmetry region, whereas outside its limits there always exists a range, centered at $Q=0$, for real propagation eigenvalues.  The symmetry breaking transitions in this case are found at $k_1=0.\bar{1}$ and $k_2=9$.  If $k$ is below $k_1$ or exceeds $k_2$ real eigenvalues can no longer be obtained.

\subsection{Propagation properties in type II structures}

\begin{figure}
\centering
\includegraphics[width=8cm]{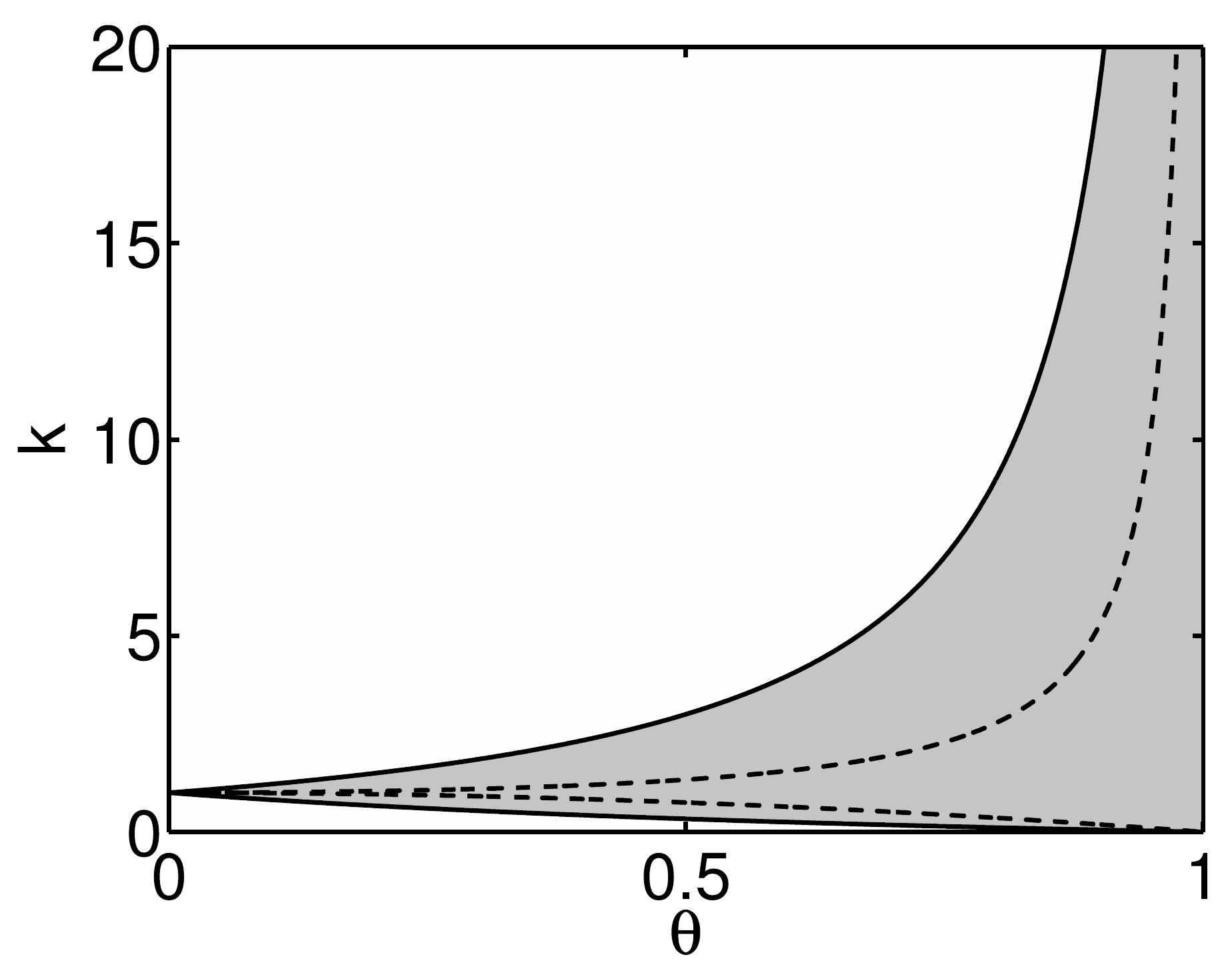}
\caption{Solid lines are the boundaries for the $\cal{PT}$ symmetry thresholds and dashed lines the limits for the single-ring lasing instabilities.}\label{fig::regions}
\end{figure}

\begin{figure}
\centering
\begin{tabular}{cc}
(a)&(b)\\
\includegraphics[width=.45\columnwidth]{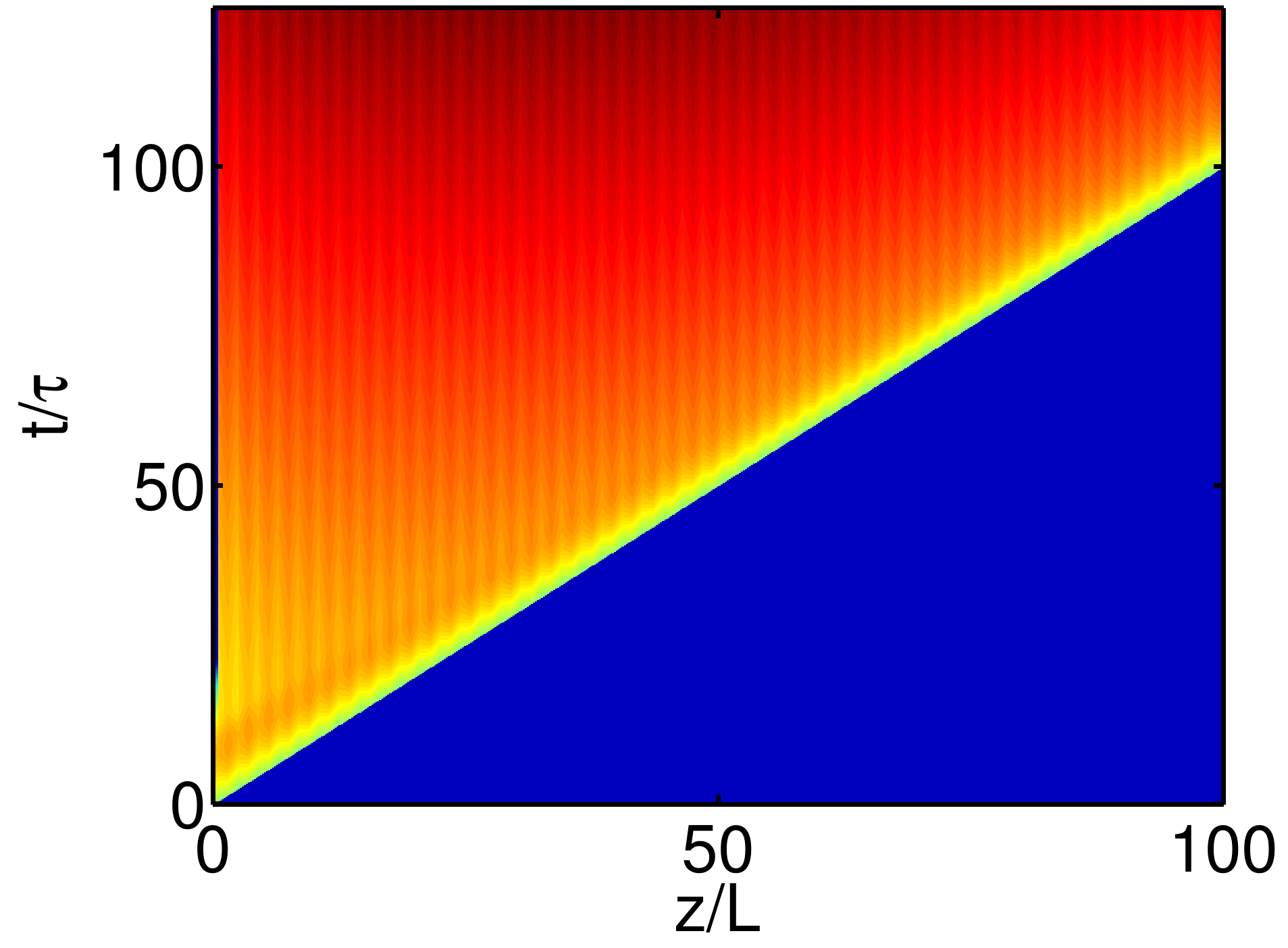}&\includegraphics[width=.45\columnwidth]{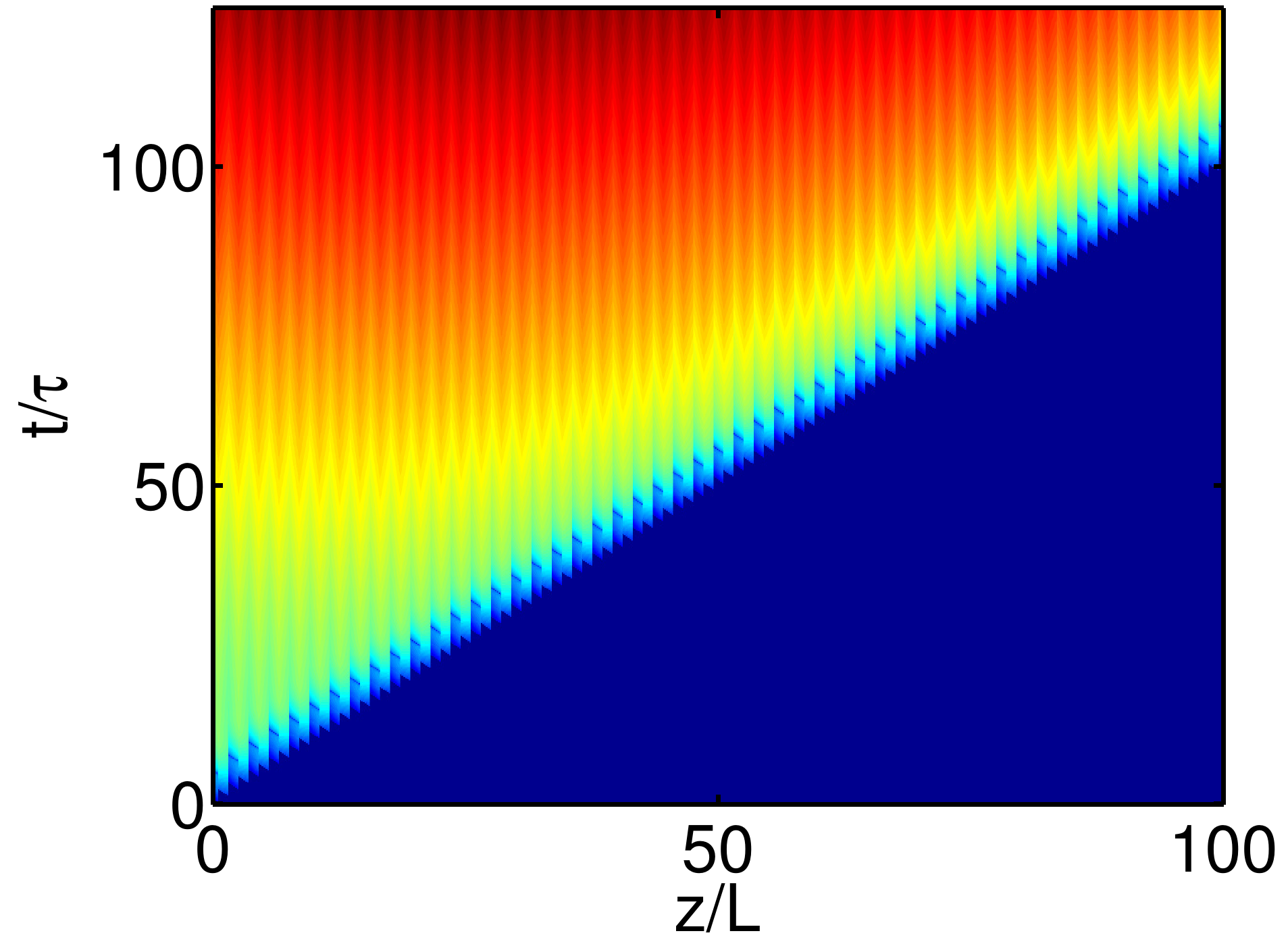}\\
(c)&(d)\\
\includegraphics[width=.45\columnwidth]{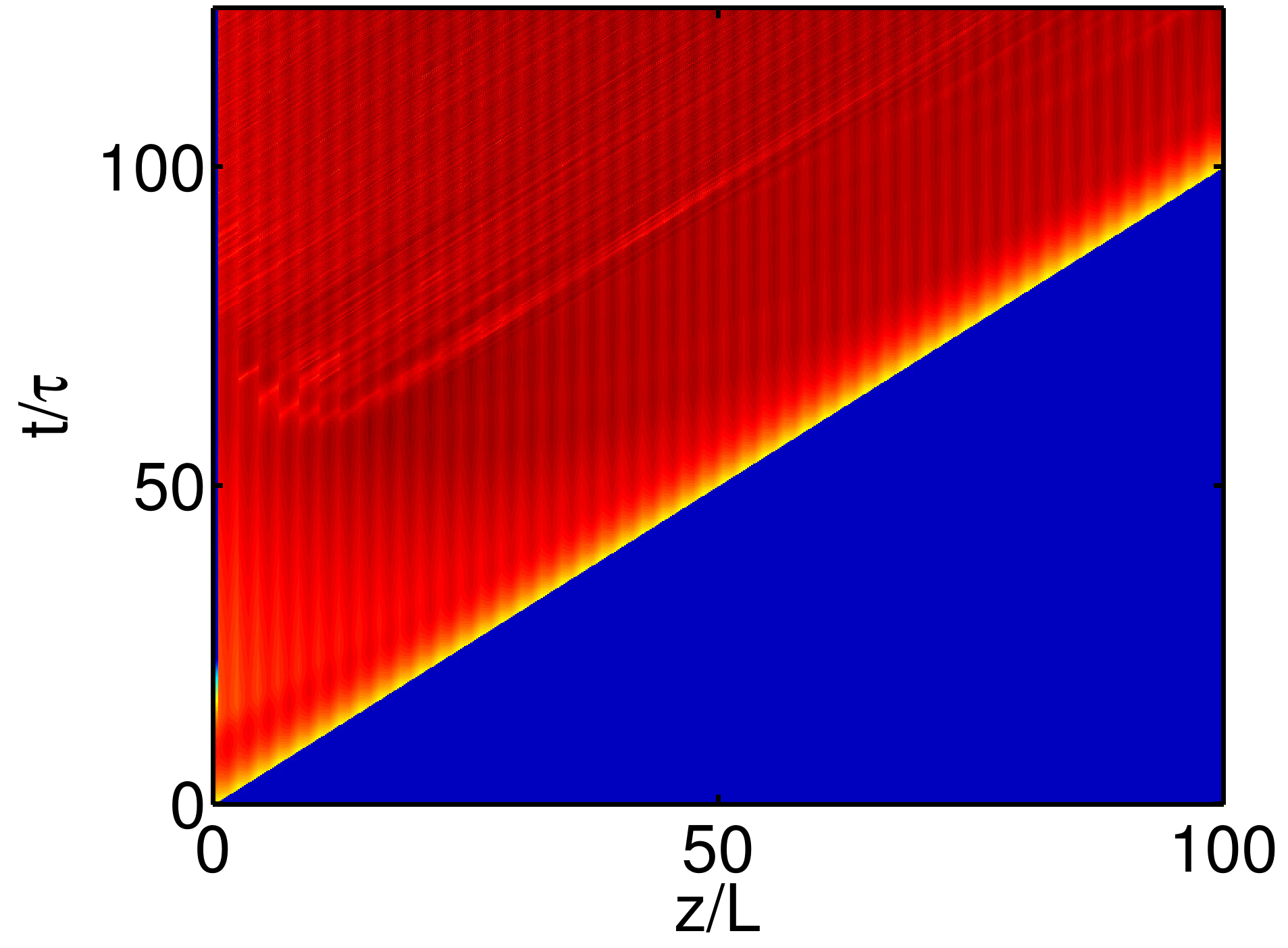}&\includegraphics[width=.45\columnwidth]{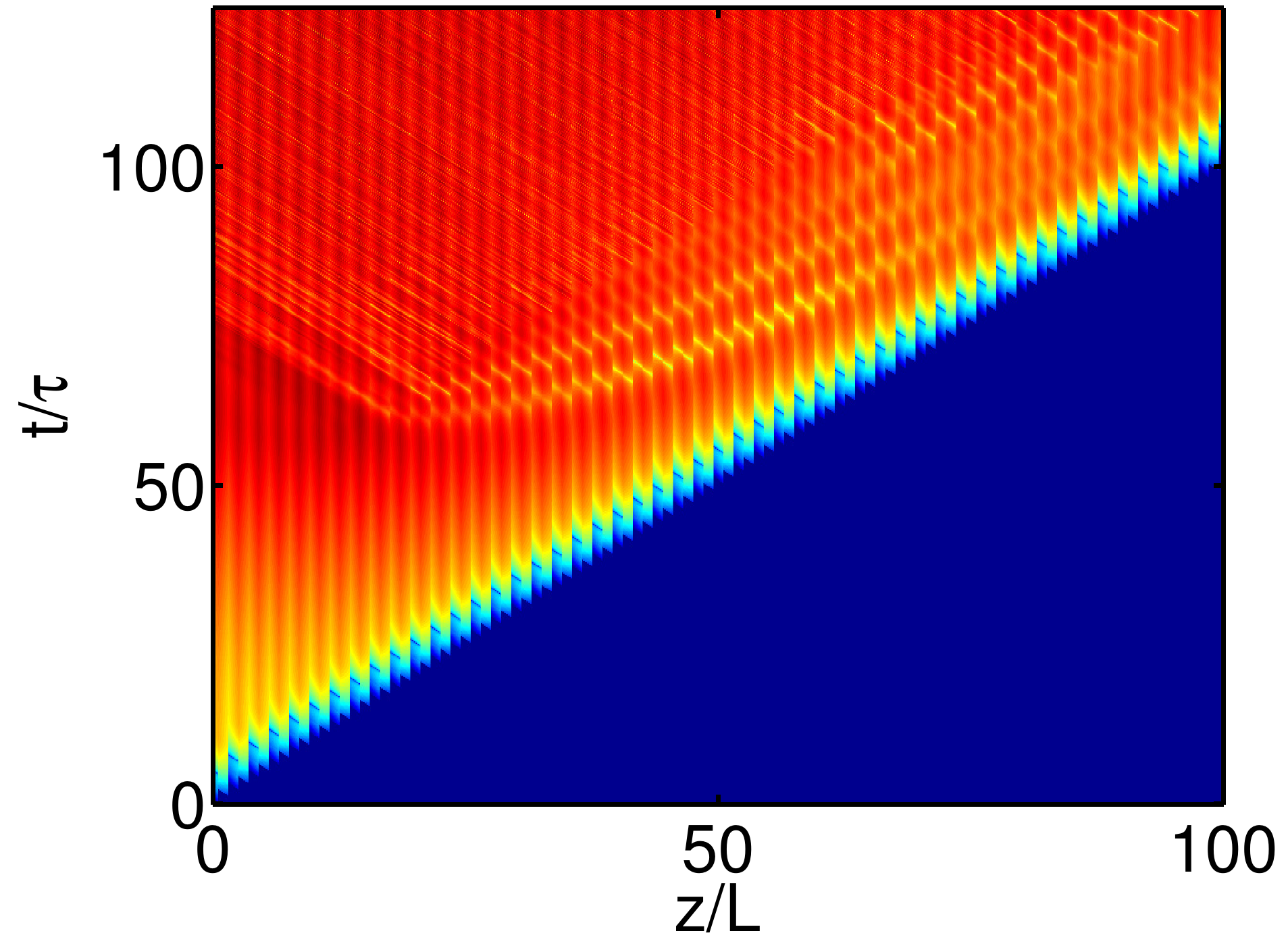}\\
\end{tabular}
\caption{Logarithmic plot of the field amplitude in a $|\theta|=0.995$ CROW at $k=0.01$, close to the gain threshold for individual ring lasing. (a) and (b) are the $A$ and $B$ components, respectively, in the linear case and (c) and (d) in the presence of a small nonlinearity ($\Gamma=0.001$).} \label{fig:sb}
\end{figure}

\begin{figure}
\centering
\begin{tabular}{cc}
(a)&(b)\\
\includegraphics[width=.45\columnwidth]{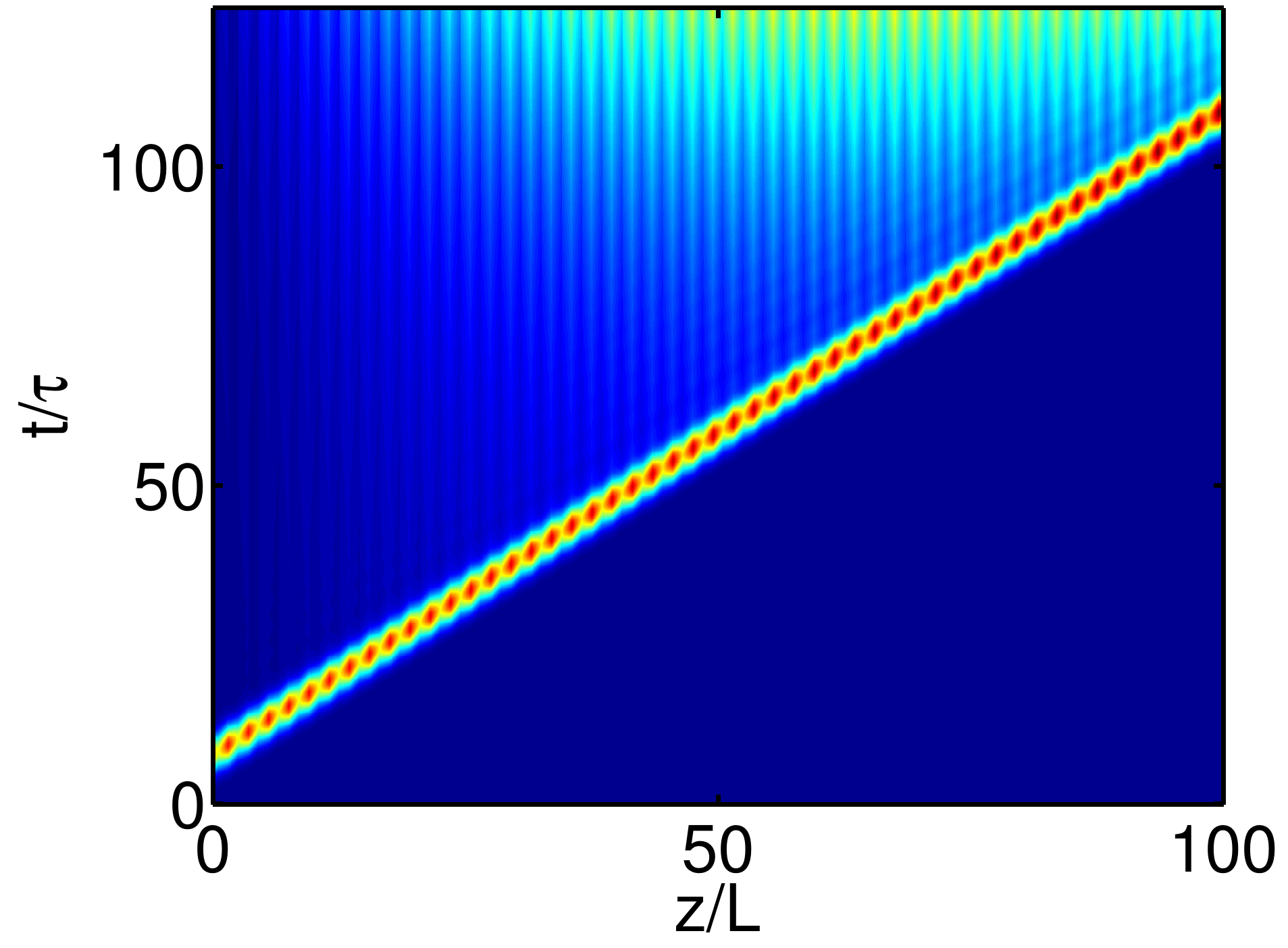}&\includegraphics[width=.45\columnwidth]{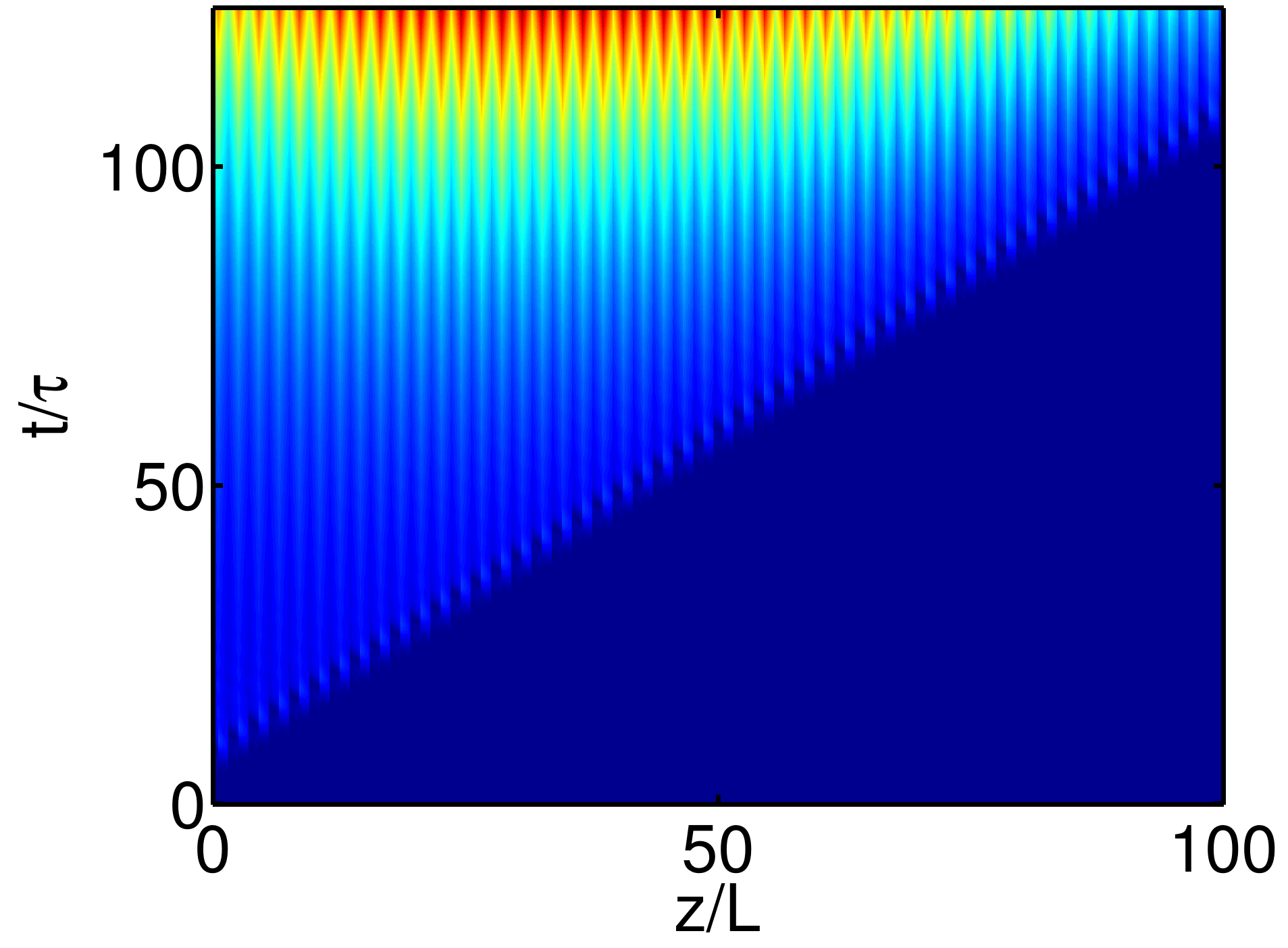}\\
(c)&(d)\\
\includegraphics[width=.45\columnwidth]{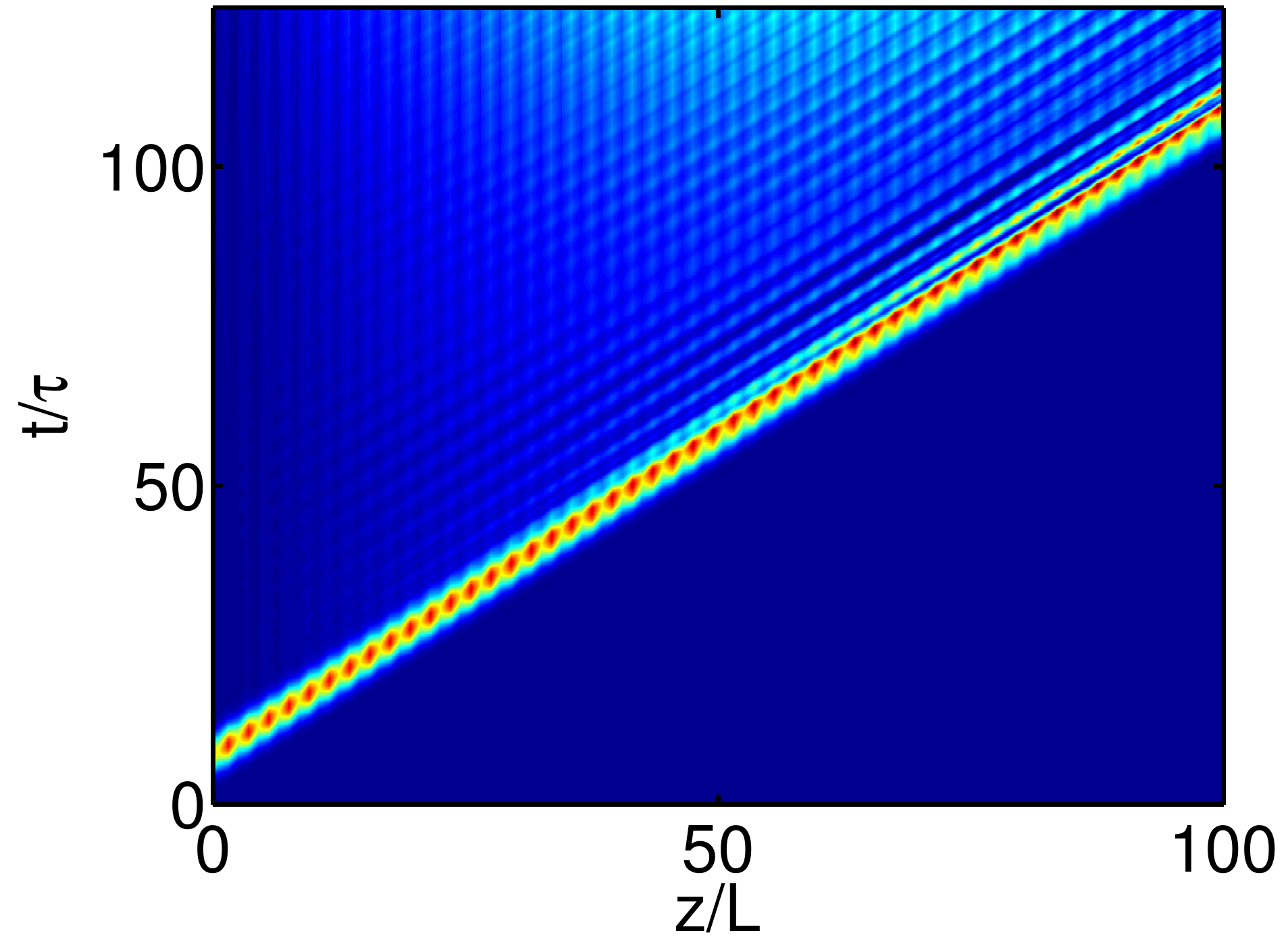}&\includegraphics[width=.45\columnwidth]{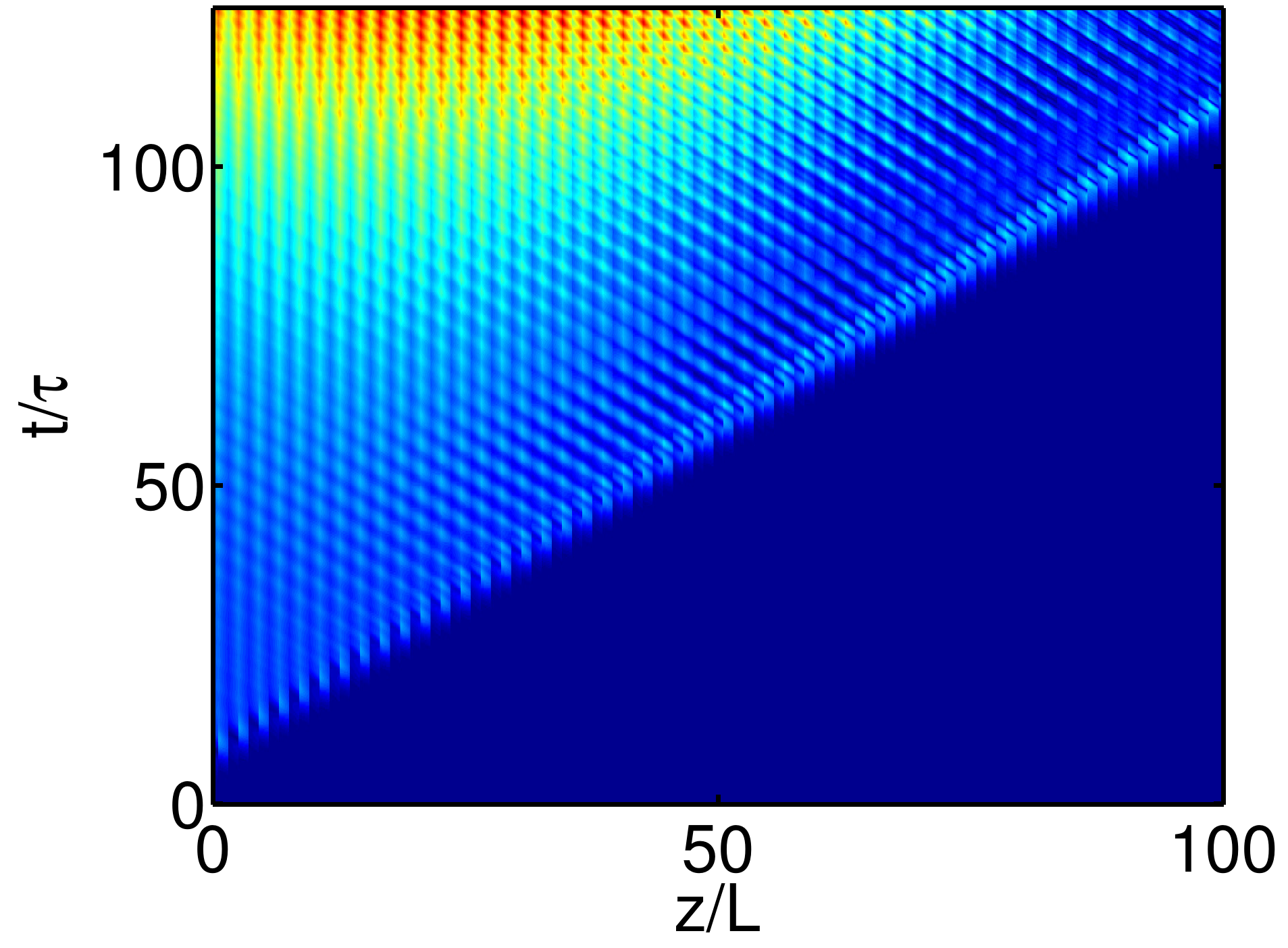}\\
(e)&(f)\\
\includegraphics[width=.45\columnwidth]{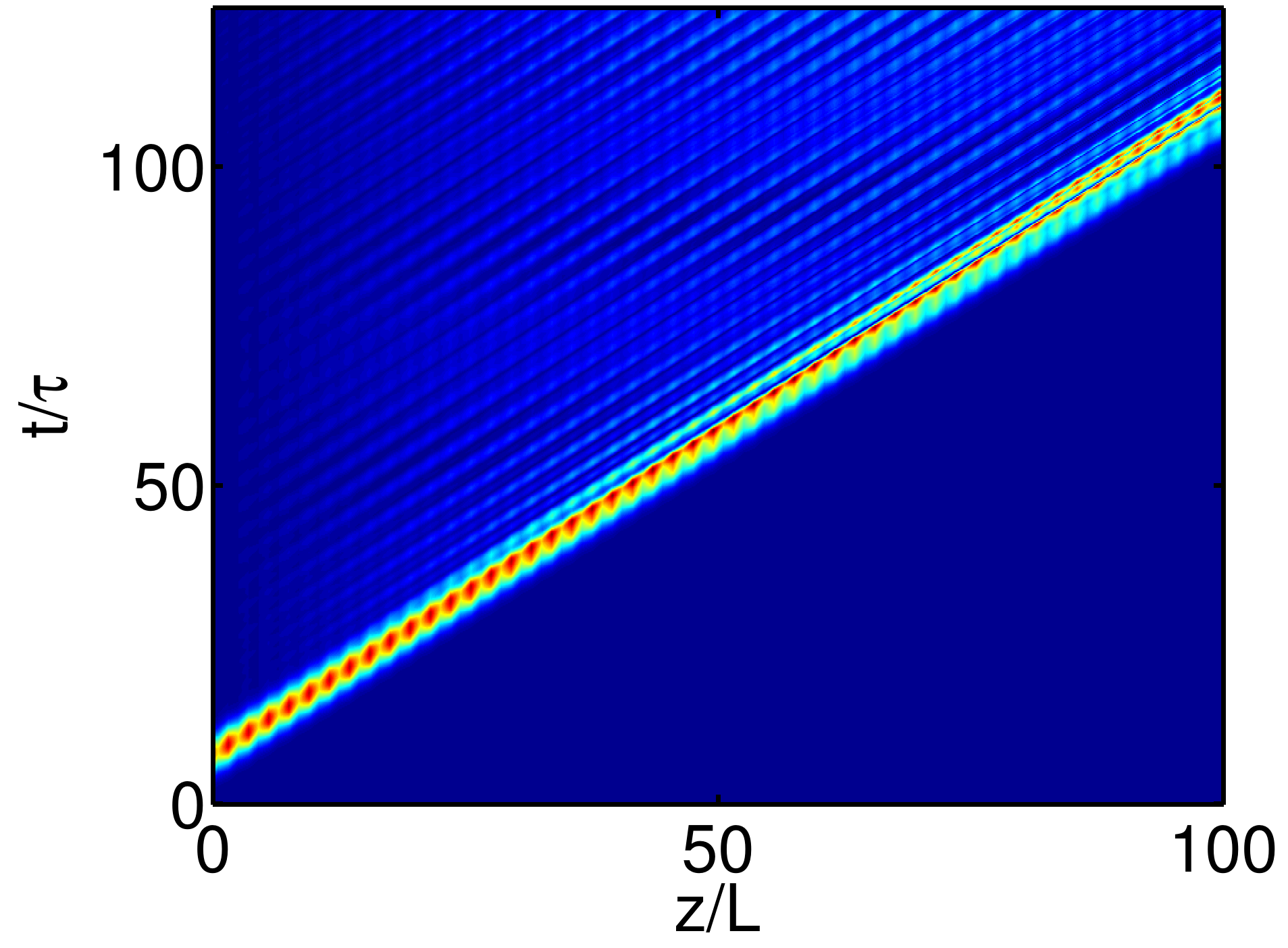}&\includegraphics[width=.45\columnwidth]{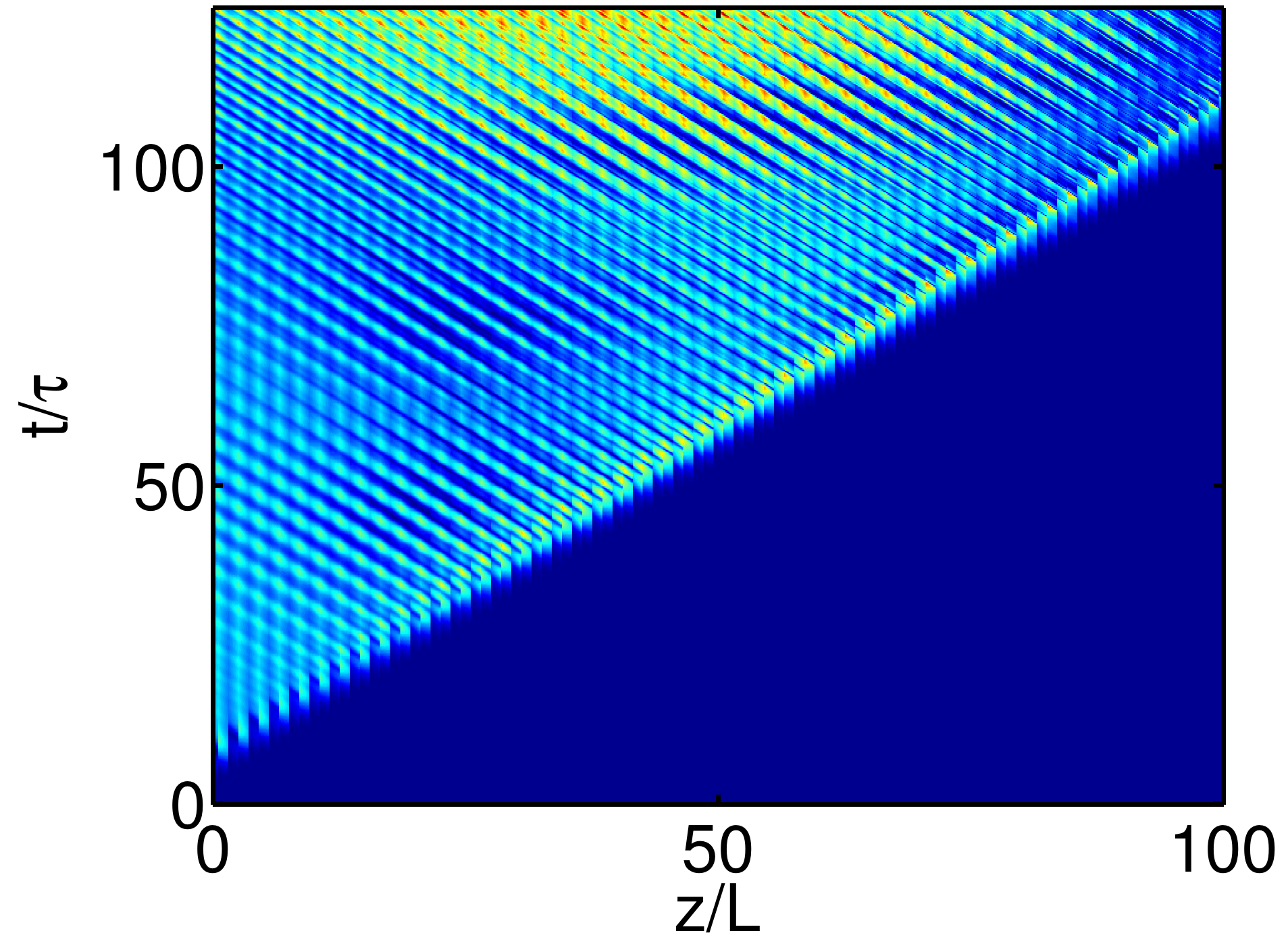}\\
\end{tabular}
\caption{Plot of the field amplitude for a pulse input in a tightly coupled $|\theta|=0.995$ CROW at $k=0.25$. (a) and (b) are the $A$ and $B$ components, respectively, in the linear case, (c) and (d) for a nonlinearity parameter $\Gamma=0.025$ and (e) and (f) for $\Gamma=0.05$ .} \label{fig:nsb}
\end{figure}

\begin{figure}
\centering
\begin{tabular}{cc}
(a)&(b)\\
\includegraphics[width=.45\columnwidth]{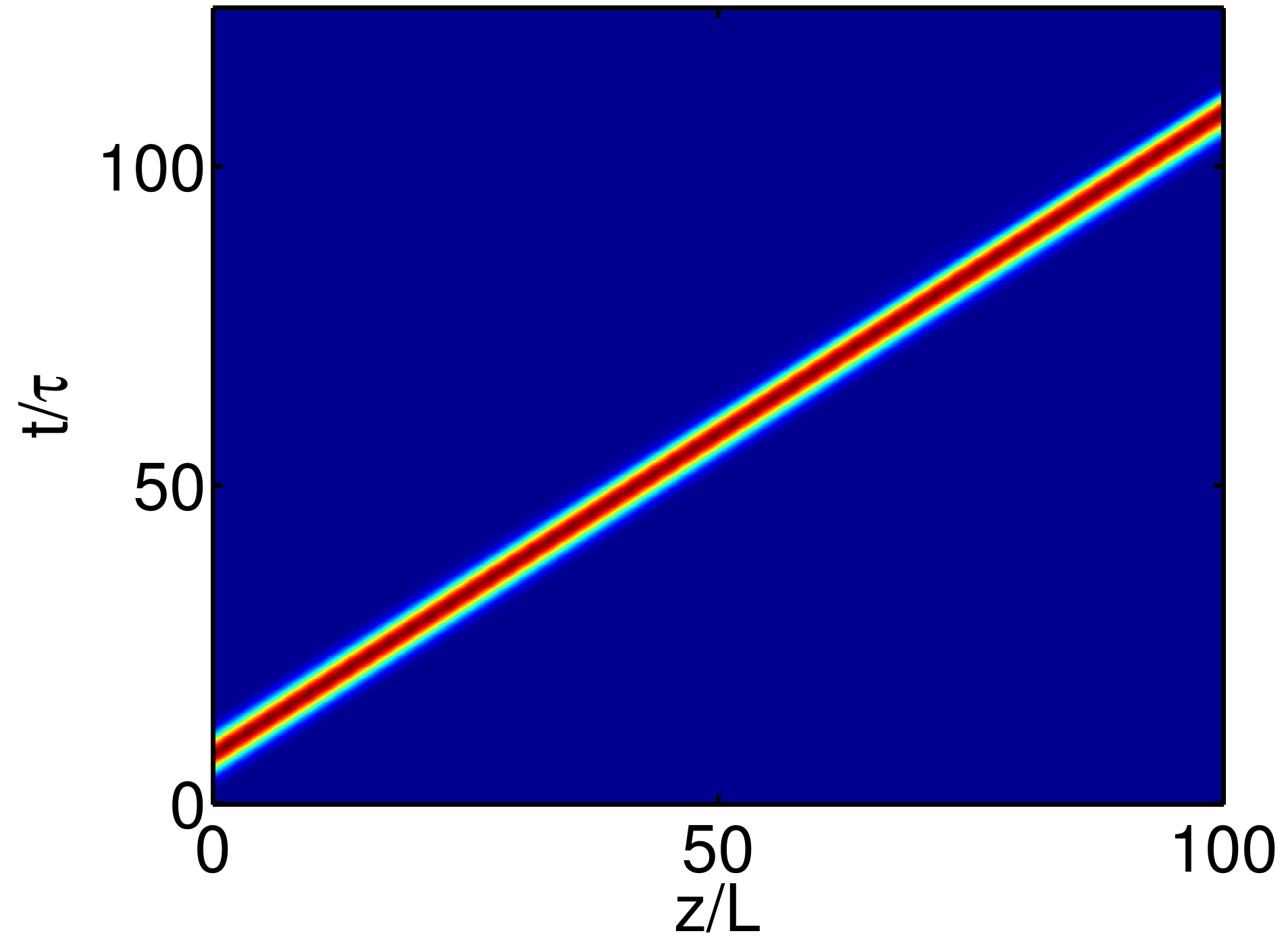}&\includegraphics[width=.45\columnwidth]{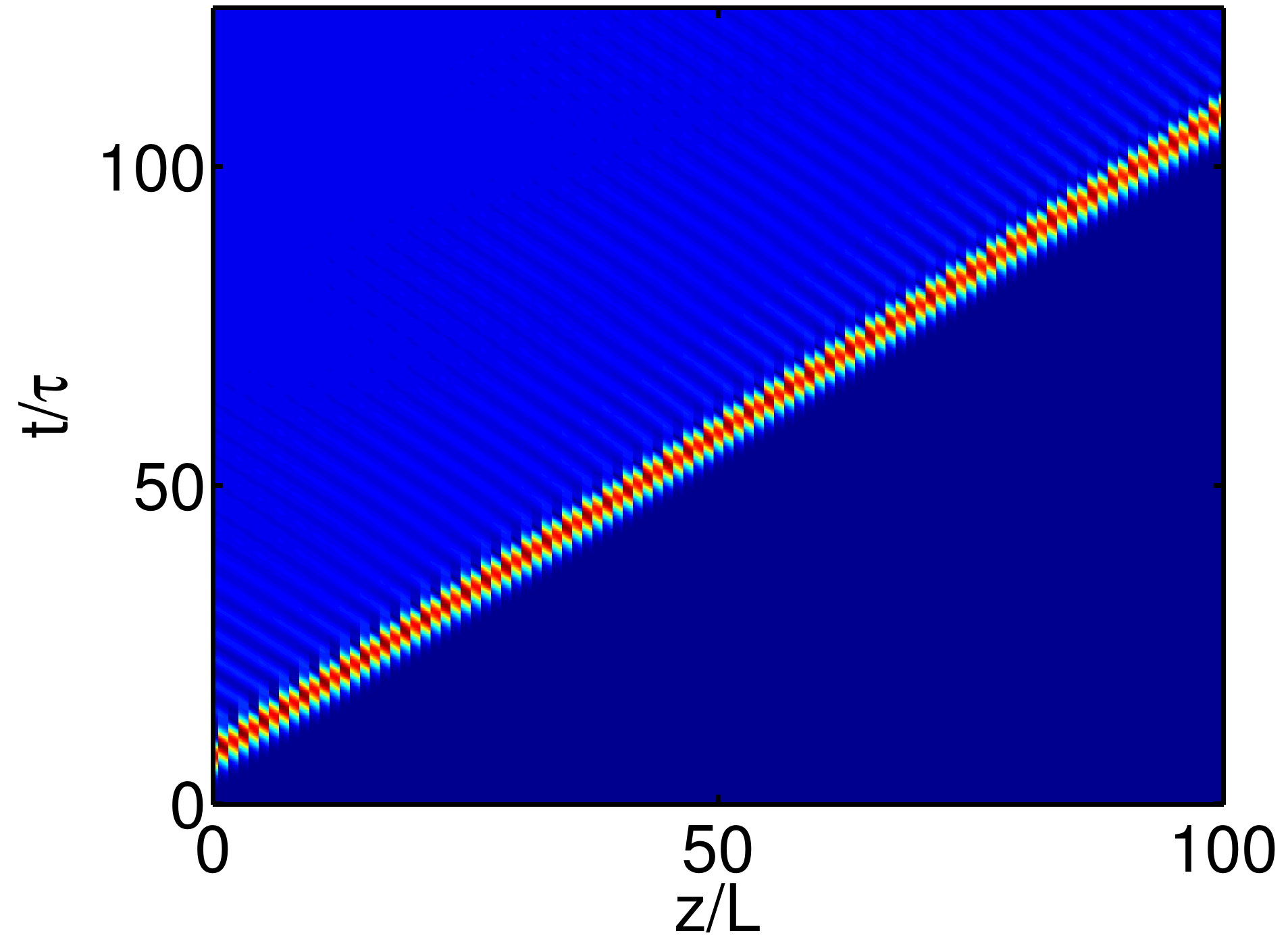}\\
(c)&(d)\\
\includegraphics[width=.45\columnwidth]{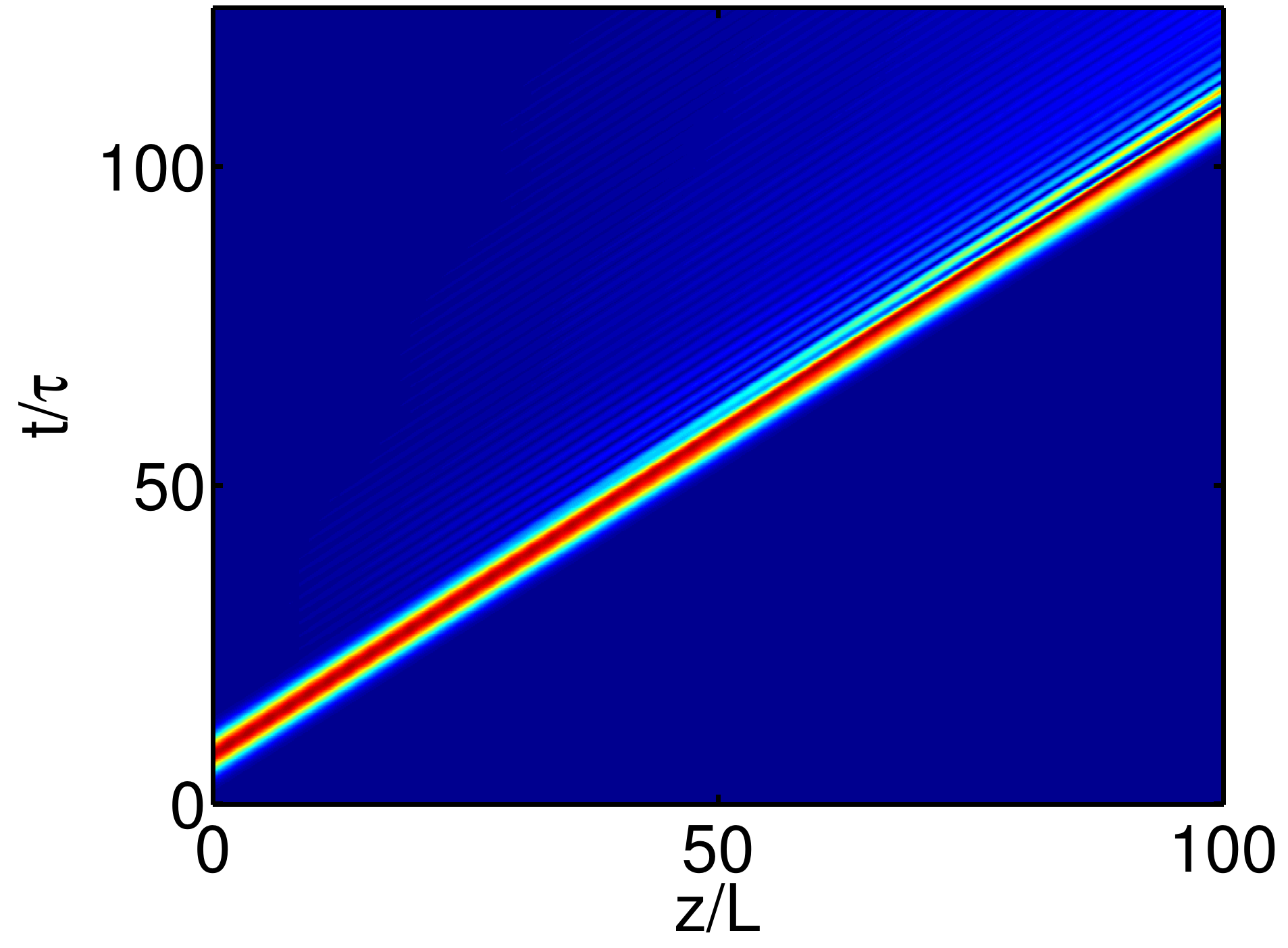}&\includegraphics[width=.45\columnwidth]{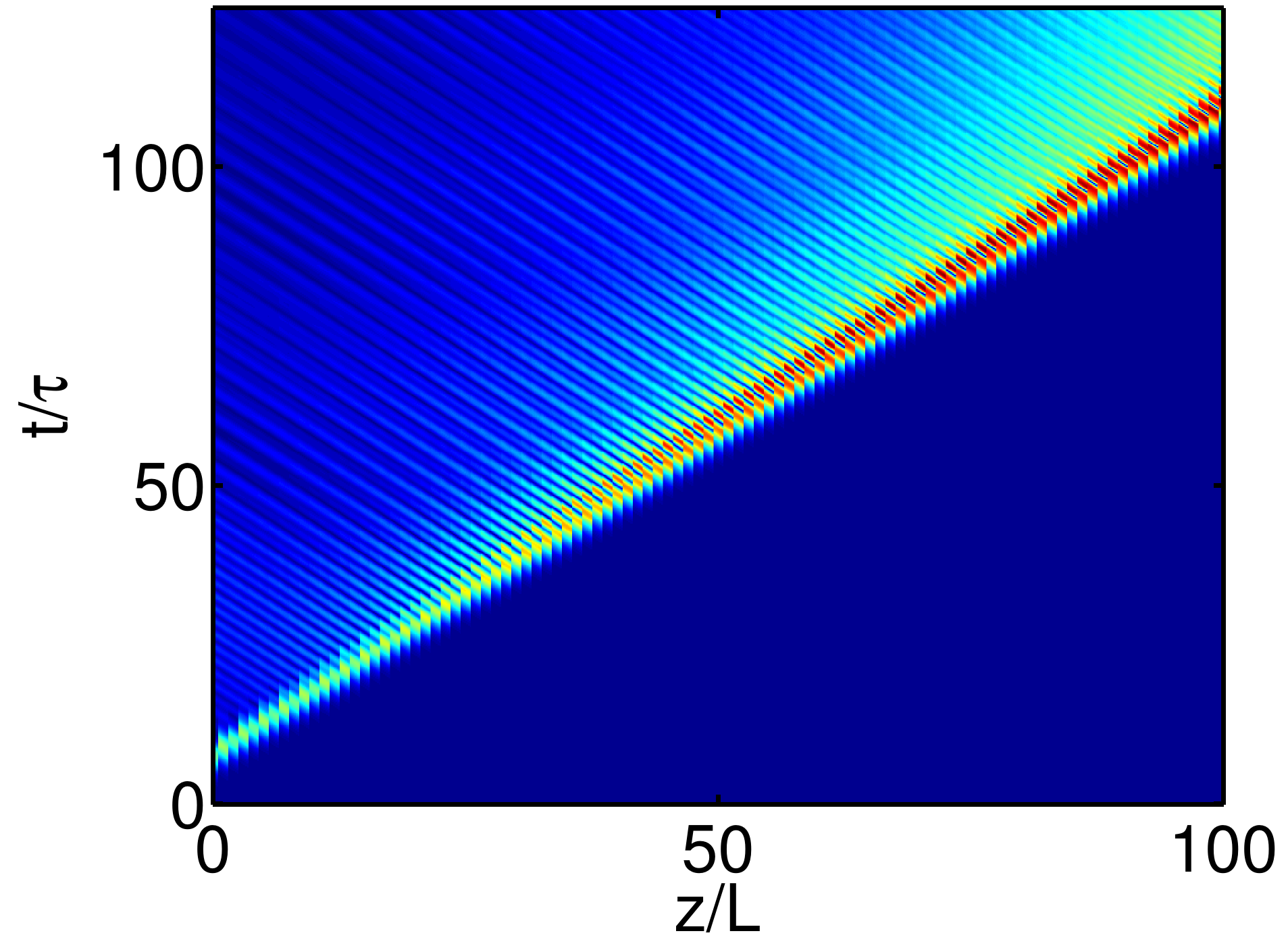}\\
(e)&(f)\\
\includegraphics[width=.45\columnwidth]{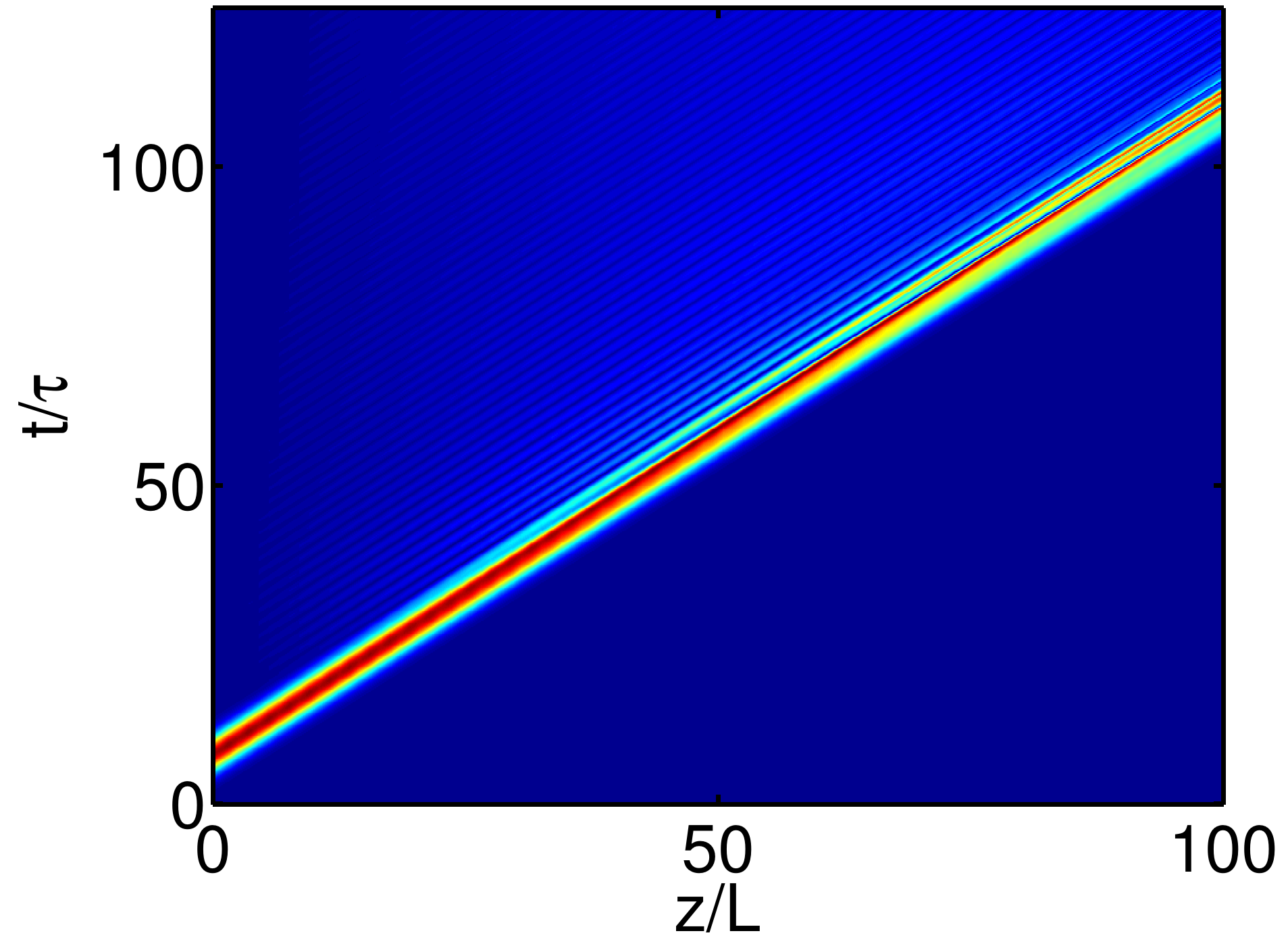}&\includegraphics[width=.45\columnwidth]{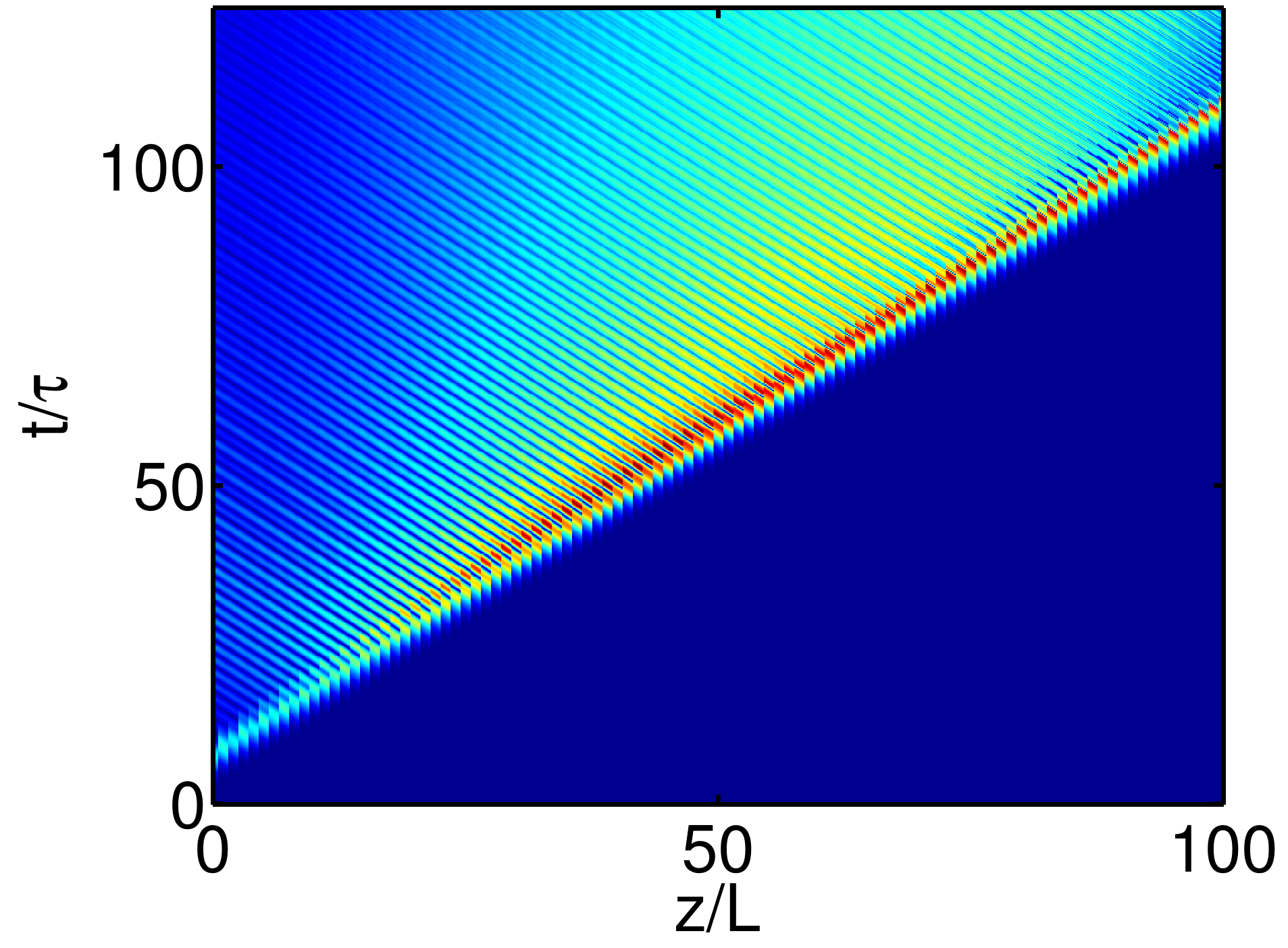}\\
\end{tabular}
\caption{Plot of the field amplitude in a tightly coupled $|\theta|=0.995$ CROW at $k=0.9025$, close to the stable limit value of $k=1$, for a pulse input. (a) and (b) are the $A$ and $B$ components, respectively, in the linear case, (c) and (d) for a nonlinearity parameter $\Gamma=0.025$ and (e) and (f) for $\Gamma=0.05$ .} \label{fig:casiestable}

\end{figure}

\begin{figure}
\centering
\begin{tabular}{cc}
(a)&(b)\\
\includegraphics[width=.45\columnwidth]{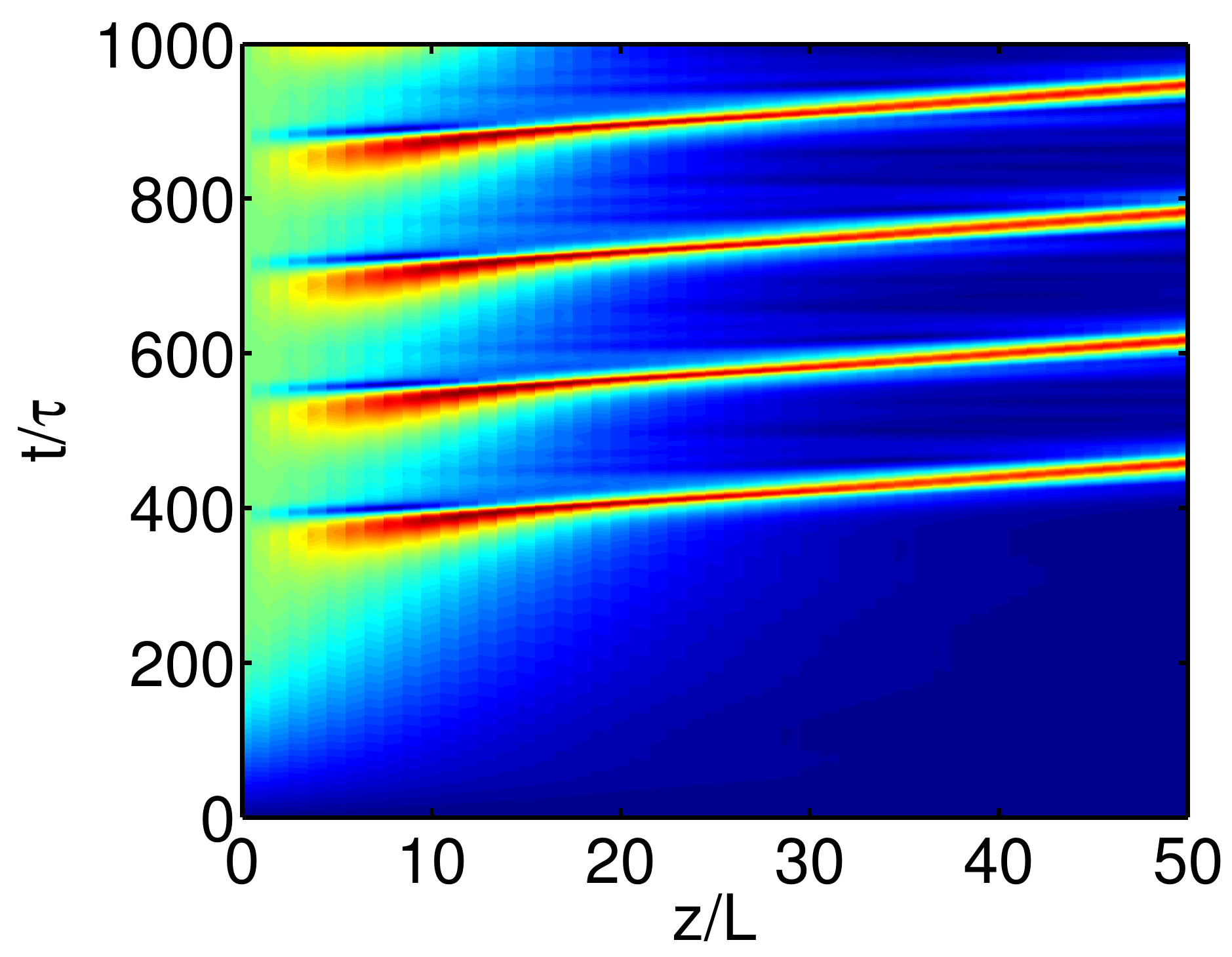}&\includegraphics[width=.45\columnwidth]{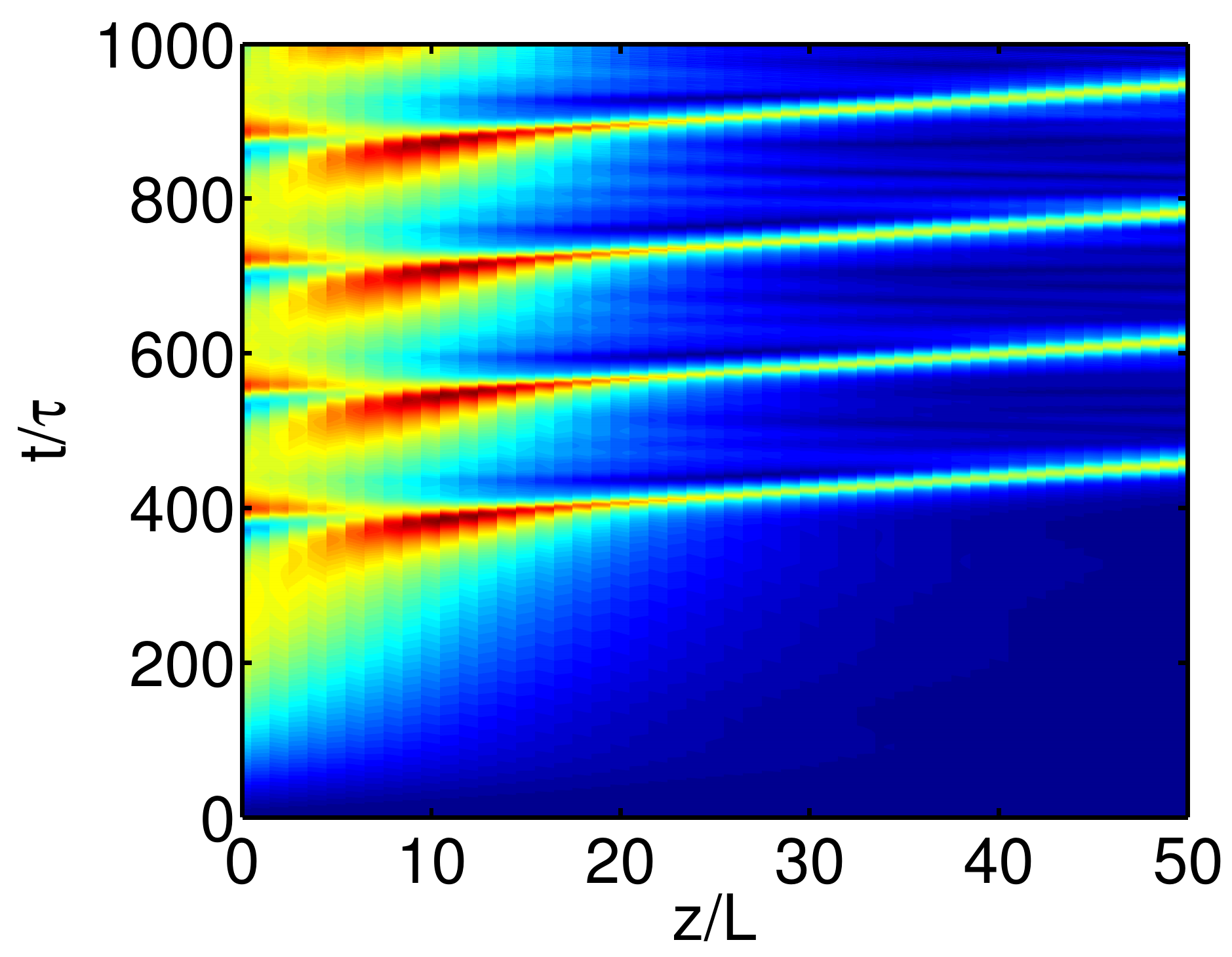}
\end{tabular}
\caption{Intracavity field in a type II structure in the quasi stable regime with $|\theta|=0.992$, $k=0.9025$ and $\Gamma=0.055$ for a step input. (a) is the $A$ component of the field and (b) is the $B$ component. } \label{fig:gapII}
\end{figure}

One basic consideration regarding the stability of the propagating solutions is the lasing threshold of an individual microring.  The total loop gain is given by $g_l^2a_l^2(1-\theta^2)$, where $l=1,2$ and $\rho=\sqrt{1-\theta^2}$ is the loss factor at each coupler. The lasing stability region is, therefore, defined by the two conditions $g_l^2a_l^2\rho^2 \le 1$, $l=1,2$, which are equivalent to the inequality
\begin{equation}
{1-\theta^2}\le k \le \dfrac{1}{{1-\theta^2}}.\label{eq::lasing}
\end{equation}

Outside this region, we expect an intrinsic instability arising from the build-up of an exponentially growing oscillation at the resonance frequency, since no saturation is present in the system.  It is important to note that the effect of saturation and, more generally, the gain dynamics \cite{ol2009} can play a key role in the stabilization of active coupled microresonator chains, and the study of their effect will be the subject of a subsequent work.

The solutions are expected to be unstable outside the limits of the region set by the condition \eqref{eq::lasing}.  Lasing unstable modes correspond to the set up an ever-growing oscillation at each microring and the build up shows up simultaneously  in the whole structure.  Figure \ref{fig::regions} shows the parity-time symmetry breaking thresholds \eqref{eq::pt} with solid lines and the lasing instability thresholds \eqref{eq::lasing} with dashed lines. Therefore unstable soutions are expected even outside the broken symmetry parameter domain.  The origin of this instability has been described from the simple physical grounds of the existence of a lasing oscillation at individual microrings, but does not exclude an even broader instability domain since the potential laser oscillators are coupled together.  In fact, the dispersion equation has bands with imaginary $\Omega$ centered at $Q=\pi$ for all values of $k$ except at $k=1$.  Figure \ref{fig::ptbands} show these imaginary $\Omega$ bands as dashed lines.   

It can be observed in Figure \ref{fig::regions} that as $|\theta|\to 1$, in the tightly coupling regime, the domain of values of $k$ between the symmetry breaking transitions becomes broader.  Similarly,  the instability regions corresponding to the existence of an imaginary part in $\Omega$ at the edges of the first Brillouin zone become smaller (see Figure \ref{fig::ptbands}) and quasi-stable propagation can be observed close to the stable limit value of $k=1$ in tightly coupled CROWs.

 The evolution of an input Gaussian pulse in various tightly coupled CROW structures with $|\theta|=0.995$ is shown in Figures \ref{fig:sb}, \ref{fig:nsb} and \ref{fig:casiestable}.  In all cases, propagation takes place within the boundaries set by condition  \eqref{eq::lasing}.   Figure \ref{fig:sb} displays the space-time dynamics of an optical field at $k=0.01$, very close to the onset of the lasing instability for an individual rings.  In this case, an unstable non-propagating optical field grows very fast and a logarithmic plot has been employed.  The presence of a nonlinear response creates a more complex field structure when a certain intracavity field level is exceeded.  This is illustrated in Figures \ref{fig:sb} (c) and (d) for a small nonlinearity parameter.

The propagation of the same input pulse at $k=0.25$, closer to the stability limit $k=1$ is displayed in Figure \ref{fig:nsb}.  In the linear case,  a nonpropagating instability starts to develop with a growth rate much smaller than in the former case.  This component is again due to the existence of unstable bands centered at the edges of the first Brillouin zone.  The presence of nonlinearity produces an interesting modification of the propagation properties of the optical field:  The relative amplitude of the backward $B$ component of the leading pulse grows with the nonlinearity, the growth rate of the unstable component is reduced and multiple peaks in the main pulse start to develop as the nonlinearity grows.

When the propagation takes place close enough to the stable propagation limit $k=1$, in the linear case, a quasi-stable  pulse without evidence of instability growth for the time duration of the simulation is obtained.  As the nonlinearity in the system is increased, an emission from this main pulse and the structuring of the pulse itself are observed.  This is shown in Figures  \ref{fig:casiestable} (c) to (f), where Figures \ref{fig:casiestable} (a) and (b) correspond to the linear case.  

Figure \ref{fig:gapII} displays the spontaneous generation of a train of solitons from an input CW step signal in the quasi stable regime described above.  This process is very similar to that of the structure without gain and loss.

\section{Discussion}

The propagation of optical fields in parity-time symmetric linear and nonlinear microring CROWs has been studied.  Two different types of microring structures with periodic gain and loss have been considered.  When the gain-loss period coincides with the microring pitch,  no symmetry breaking transition is observed and the main effects due to the presence of gain and loss are an effective modification of the nonlinearity parameter and an alteration of the properties of the nonlinear Bloch modes in the bandgap that permits to keep the balance of the forward/backward components.  The spontaneous generation of gap solitons in these structures has been demonstrated through numerical simulation.

More general structures with a gain/loss period that is double of the microring sequence have also been considered.  The characteristics of the band structure in the linear case have been shown to depend on an gain/loss parameter $k$ and the coupling coefficient $\theta$ between neighbouring resonators.  At two specific values of $k$ for each $\theta$ there exist symmetry breaking transitions.  By crossing these treshold values of $k$,  the eigenvalues defining the propagation properties become complex and rapidly growing instabilities are found in the field evolution.  Outside the broken symmetry region, stable propagation is strictly found only at $k=1$, but quasi-stable propagation is obtained in the tightly coupling regime when $k$ is sufficiently close to $1$.   The propagation of an input pulse is affected by the presence of nonlinearity in the structure, which produces the shedding of radiation and the modification of the pulse structure during propagation.  When larger deviations from the $k=1$ value are addressed, the growth of the unstable solutions is also observed.  The growth rate can be controlled with the level of the nonlinearity.

\section{Acknowledgments}
This work has been funded by the Spanish Ministerio de Educación y Ciencia and FEDER, project number TEC2010-21303-C04-04 and Junta de Castilla y Le\'on Project No. VA300A12-1.

\end{document}